\documentclass[aps,pra,reprint,amsmath,amssymb,superscriptaddress,onecolumn,longbibliography,notitlepage,nofootinbib]{revtex4-2}
\usepackage{mathtools}
\usepackage{siunitx}
\usepackage[hidelinks]{hyperref}
\usepackage{svg}
\usepackage{caption}
\usepackage{float}
\usepackage{subcaption}
\usepackage[braket, qm]{qcircuit}
\usepackage{bbm}

\usepackage{soul}
\usepackage{siunitx}
\usepackage{wrapfig}
\usepackage{caption}

\usepackage{enumitem} 
\usepackage{tikz}
\newcommand*\circled[1]{\tikz[baseline=(char.base)]{
            \node[shape=circle,draw,inner sep=2pt] (char) {#1};}}

\newcommand{\unsim}{\mathord{\sim}} 

\usepackage{xcolor}

\begin{document}
\title{The physics of optical computing}
\author{Peter~L.~McMahon}
\email{pmcmahon@cornell.edu}
\affiliation{School of Applied and Engineering Physics, Cornell University, Ithaca, NY 14853, USA}

\begin{abstract}
There has been a resurgence of interest in optical computing over the past decade, both in academia and in industry, with much of the excitement centered around special-purpose optical computers for neural-network processing. Optical computing has been a topic of periodic study for over 50 years, including for neural networks three decades ago, and a wide variety of optical-computing schemes and architectures have been proposed. In this paper we provide a systematic explanation of why and how optics might be able to give speed or energy-efficiency benefits over electronics for computing, enumerating 11 features of optics that can be harnessed when designing an optical computer. One often-mentioned motivation for optical computing---that the speed of light $c$ is fast---is \textit{not} a key differentiating physical property of optics for computing; understanding where an advantage could come from is more subtle. We discuss how gaining an advantage over state-of-the-art electronic processors will likely only be achievable by careful design that harnesses more than one of the 11 features, while avoiding a number of pitfalls that we describe.

\end{abstract}

\maketitle

\section{Introduction}
\label{sec:intro}
There has been a resurgence of interest in optical computing over the past decade, both in industry and academia \cite{wetzstein2020inference,shastri2021photonics,greengard2021photonic,mohseni2022ising}. What is the fundamental physical basis upon which we can expect an optical computer to outperform an electronic computer, at least for some tasks? In this Perspectives piece we enumerate and discuss 11 features of optics and optical computing that can contribute to an advantage for an optical computer. Any optical computer that achieves an advantage in practice will likely need to harness more than one of these features. An explicit list of features can help to make clear what ingredients the architect of an optical computer has to work with. It also allows us to systematically identify the fundamental physical principles behind the operation of different proposed optical computers, aid us in analyzing what advantage they can hope to achieve, and how their designs might be improved by exploiting further features. The design of a successful optical computer must be carefully engineered to avoid bottlenecks or overhead that would outweigh the optical benefits. We discuss some of the pitfalls and approaches one can take to mitigate them.

The high bar set by electronic processors has contributed to periods when there has been pessimism about the prospects for optical computing (for example, see Refs.~\cite{tucker2010role,miller2010optical} from just over a decade ago). Given the continued improvements in complementary metal--oxide--semiconductor (CMOS) technology \cite{datta2022toward}, \textbf{why is there now renewed excitement about optical computing?}\footnote{As some evidence that there is renewed excitement, Ref.~\cite{shen2017deep} has been cited over 1000 times in just the past 2 years. Refs.~\cite{economist2022artificial,cartlidge2023photonic} discuss some of the current commercial efforts in optical computing.} One of the major criticisms of optical computing has been that optical transistors are not competitive with their electronic counterparts. The current wave of interest in optical computing is primarily focused on optical-computer architectures that are not based on replicating digital logic with optical transistors. Instead of trying to construct general-purpose, digital computers, the community is largely targeting building special-purpose, analog computers. Both these shifts---to \textit{special-purpose}, and to \textit{analog} processing---are important. Trying to build performant general-purpose processors with optics remains out of reach\footnote{This is essentially because general-purpose processors are expected to have no errors (accountants want sums in their spreadsheets to be exactly correct, for example), and we only know how to achieve error-free machines with digital logic, and to build digital logic, we need an optical transistor satisfying the criteria given in Ref.~\cite{miller2010optical}, or something similar.}, but one can alternatively build optical processors that are specialized to particular applications for which completely error-free operation is not necessary.

There are several application areas being targeted by special-purpose optical computers presently, including: (i) neural networks \cite{wetzstein2020inference}; (ii) scientific computing \cite{feinberg2018enabling}; (iii) combinatorial optimization \cite{mohseni2022ising}; (iv) cryptography \cite{dubrovsky2020towards,pai2023experimental,cartlidge2023photonic}. Matrix-vector multiplications are a key algorithmic primitive in all four application areas and are the target of much of the current research in optical computing. Fourier transforms and convolutions have applicability across neural networks, scientific computing, and cryptography, contributing to their prominence in current research.\footnote{Optical correlators have been released as commercial products during several periods over the past few decades \cite{ambs2010optical}, so this is not a new direction even commercially, but one that has been revitalized.} There is also a substantial thrust in performing computations for neural networks that are not explicitly engineered to be matrix-vector multiplications or convolutions \cite{van2017advances,wetzstein2020inference,nakajima2021scalable,teugin2021scalable,wright2022deep,zhou2022nonlinear}. A commonality among all four application areas is that the subroutines performed optically are still useful even if they suffer from some error (noise). This is crucial since it is difficult to achieve an effective precision greater than 10 bits in any analog computer, including analog optical computers, so applications of analog optical computers should be robust to this level of noise. Neural networks are a particularly good match because, at least during inference (as opposed to training), neural networks do not suffer a substantial decrease in accuracy even if they are restricted to integer arithmetic with fewer than 8 bits of precision \cite{wetzstein2020inference,semenova2022understanding}.\footnote{A concern for any analog neural-network processor, including analog optical processors, is the potential for accumulation of errors in executing deep neural networks. This has recently been theoretically analyzed, with a conclusion that deleterious effects of noise accumulation can be mitigated, even in the case of correlated noise \cite{huang2022prospects}. Uncorrelated noise that merely leads to an effective low-bit-precision has been shown in simulations of deep optical neural networks (having 60 optically executed layers) to yield accuracies that are the same as or better than that of digital electronic processors executing the same neural network with 8-bit integer arithmetic \cite{anderson2023optical}, i.e., the simulations predicted that the accumulation of error in an optical implementation of the neural network would not have a noticeable impact on accuracy versus a standard digital electronic implementation. However, for all applications of analog optical processors---not just neural networks---intuition and simulations about resilience to noise ultimately need to be validated by optical experiments.}

With this context, we can now give a fuller answer to why there is renewed excitement in optical computing:\footnote{More directly and colloquially, one can ask: \textit{we tried optical computing for neural networks in the 1980s and it fizzled, so why are we trying again now?} The three reasons we offer are: (i) \textit{Neural networks have become important again}---people lost interest in neural networks in the 1990s and it is only over the past decade that neural networks have returned to the fore, this time stronger and more important than ever. (ii) \textit{Electronics was advancing fast enough then, but is not now}---CMOS-technology improvements, both in transistor count and in clock frequency, rapidly outpaced the developments of any alternative technology in the 1980s, whereas now CMOS electronic processors are constrained by heat dissipation and energy costs, and the anticipated improvements in CMOS technology are insufficient to satisfy the growth of neural networks. (iii) \textit{Photonics technology has improved a lot since the 1980s}---technologies spanning light generation, manipulation, and detection have all improved dramatically.} (i) \textit{The rise of neural networks}---over the past decade, neural networks have become a dominant approach in machine learning and have become extremely compute-resource-intensive. This has led to strong interest in alternative hardware approaches specialized to neural networks, and the intrinsic resilience of neural networks to noise makes them well-suited to analog optical implementations. (ii) \textit{CMOS improvements won't be enough to satisfy application demand}---while there has been remarkable progress in CMOS hardware \cite{datta2022toward}, it is also simultaneously true that both for neural networks and for some other applications (such as combinatorial optimization), the anticipated future improvements in CMOS hardware \cite{leiserson2020there} are less than users would like and will limit application capabilities \cite{horowitz2014computings}.\footnote{The number of parameters in neural networks---one measure of their size and computational demand---have been growing much faster than hardware improvements \cite{xu2018scaling}, primarily because of the finding that increased scale often leads to increased capability or accuracy \cite{kaplan2020scaling,zhai2022scaling}.} (iii) \textit{Improvements in photonics hardware}---driven largely by the consumer-electronics and the optical-communications industries, there have been enormous advances in the scale, speed, and energy efficiency of photonic devices over the past 30 years since the last big surge of interest in optical neural networks.\footnote{As examples, Samsung now offers a camera with 200 million pixels \cite{samsung2022samsung}, and 400-gigabit-per-second optical transceivers using on the order of 10~W of power are commercially available.} This period has also seen the development and commercialization of photonic integrated circuits \cite{fahrenkopf2019aim}, giving a miniaturized alternative to bulk optics; there have also been substantial developments in optical materials and devices \cite{chen2011device,borghi2017nonlinear,blumenthal2018silicon,gaeta2019photonic,chen2022opportunities,panuski2022full,boes2023lithium}.

A complementary trend in the electronics community (both in CMOS and beyond-CMOS technologies), which has provided further support for the development of optical computers for neural networks, has been the development of special-purpose electronic chips for neural-network processing \cite{chen2020survey}. In many cases these chips also perform analog rather than digital matrix-vector multiplications; this has led to the development of methods for training neural networks to work well on analog hardware, many of which are also applicable to analog optical neural networks. Both analog and digital electronic neural-network chips often have dataflow architectures, especially systolic-array architectures. They also often implement the concept of compute-in-memory, meaning that the physical element storing an element of a neural network's weight matrix, for example, is also the physical element in which the multiplication by that weight takes place \cite{yu2021compute}; often the stored values can only be updated slowly, but this is acceptable for neural-network inference or other scenarios where the weights will be reused many times. Systolic-array and especially compute-in-memory architectures can have a close mapping to optical processors in which information encoded in optical signals flows through processing elements, be they arrays of spatial-light-modulator pixels (e.g., Ref.~\cite{wang2022optical}), meshes of Mach-Zehnder interferometers (e.g., Ref.~\cite{shen2017deep}), crossbars of phase-change-memory cells (e.g., Ref.~\cite{feldmann2021parallel}), or networks of microring resonators (e.g., Ref.~\cite{ohno2022si}). This parallel between the architectures of analog electronic neural-network processors and analog optical neural-network processors has allowed optical-computer architects to borrow insights from the electronic-processor community. Architectural similarities also make it easier to predict how the performance of future electronic and photonic implementations are likely to compare.\footnote{Not every optical computer for neural networks is based on similar architectures to electronic neural-network processors---and there are good reasons to deviate \cite{hooker2021hardware,wright2022deep}---but in the cases where the architectures and algorithms are comparable, performance analysis is simpler because one doesn't have to disentangle the effects of different algorithms and different architectures, and can focus on the underlying physical differences: how many parallel elements are there, how fast can data be sent through them, and so on.} There are likewise architectural and algorithmic parallels between many special-purpose electronic processors for combinatorial optimization, and optical approaches for the same application area \cite{mohseni2022ising}.

In this Perspective, we limit ourselves to discussing \textit{classical} optical computing and do not review the benefits of optics for building \textit{quantum} computers \cite{rudolph2017optimistic}. We will also not attempt to compare optical classical computers with optical quantum computers, other than to say that both are competing against classical digital electronic computers but with rather different applications targeted for potential advantage \cite{hoefler2023disentangling}.

\pagebreak

\textbf{What do optical computers need to beat?} Before we discuss how an optical computer could beat an electronic computer, let's first briefly describe what they are up against and why this makes electronic processors such stiff competition. There is both a hardware and an algorithms or software component to this. On the hardware side, electronic processors based on CMOS transistors have enormous parallelism, with up to $\unsim 10^{11}$ transistors per chip, operating at a clock rate of between $\unsim \SI{1}{\giga\hertz}$ and $\unsim \SI{10}{\giga\hertz}$, and a switching energy of $<\SI{10}{\atto\joule}$ (i.e., $<10^{-17}\,\textrm{J}$) \cite{datta2022toward}. This allows modern processors to have enormous computing throughput---for example, the Nvidia H100 processor \cite{nvidia2022nvidia} can perform $4 \times 10^{15}$ 8-bit scalar multiplications per second, which corresponds to performing approximately $4 \times 10^{6}$ multiplications in parallel per clock cycle; the chip draws $<1000$~W of power. On the software side, in parallel with the $>50$ years of effort that has gone into improving transistor-based hardware, there has been $>50$ years of effort in designing algorithms\footnote{Ref.~\cite{leiserson2020there} notes that in some cases, improvements in algorithms over the past several decades have been responsible for almost as much benefit as improvements in hardware.}, and in many cases the algorithms have been implicitly or explicitly designed to be optimized for the kinds of hardware that were or are available at the time \cite{hooker2021hardware}, raising the barrier to entry for new hardware paradigms.

We will now proceed to explain what physics differences between electronics and optics can contribute to an advantage for optical computers, and then in the Discussion section we will talk about why practical advantage from optical computers has remained elusive and what paths there are to achieving advantage.

\newpage

\section{The 11 features}

Paraphrasing H.\,L.~Mencken, there is an explanation for optical computing's potential advantage that is neat, plausible, and wrong: the fact that light travels fast. We list below 11 features of either optics itself, or of a way computing can be done with optics, that are ingredients for the construction of optical computers; these features allow for explanations of how optics can deliver an advantage that are subtler but correct. We also address how the speed of light \textit{is} related to optical computing, even though it is not the cause of optical advantage.

\newpage

\begin{figure}[h]
  \includegraphics[width=0.5\textwidth]{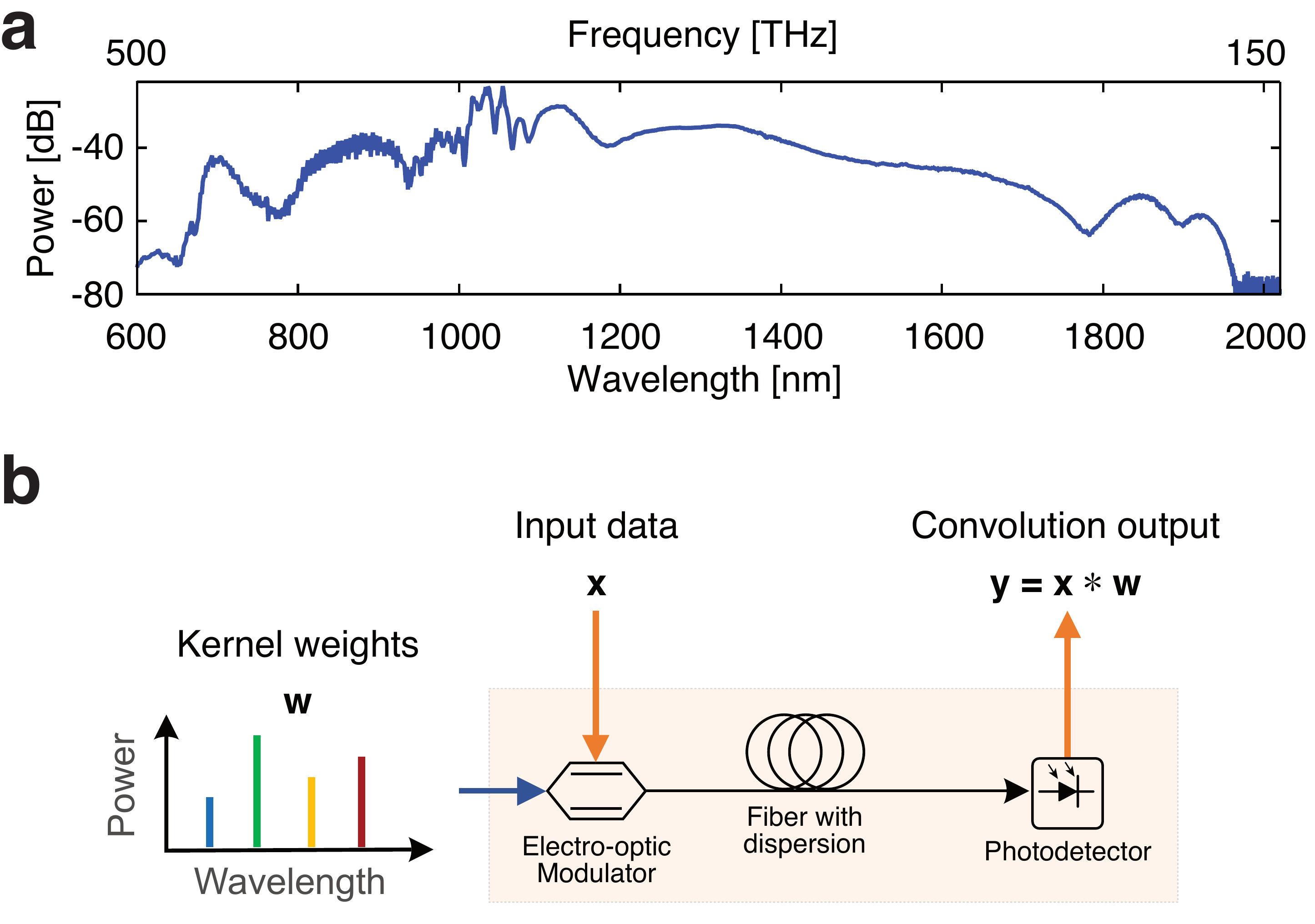}
  \caption{\label{fig:1_Bandwidth} \protect\circled{1} \textbf{Bandwidth.} \textbf{a}, An optical signal with bandwidth $>\SI{300}{\tera\hertz}$. \textbf{b}, An example of the use of frequency multiplexing in optical computing: kernel weights for a convolution are input as intensity modulations of spectral lines in a frequency comb; the use of multiple comb lines allows multiple computations to be performed in parallel. {\color{gray} Panel \textbf{a} adapted from Ref.~\cite{johnson2015octave}, \textcopyright~Optica Publishing. Panel \textbf{b} adapted from Ref.~\cite{xu2021tops}, \textcopyright~Springer Nature.}}
\end{figure}

\begin{enumerate}[label=\protect\circled{\arabic*}]

\item \textbf{Bandwidth}: photonics has a $\unsim 100,000\times$ larger bandwidth $B$ than electronics ($\unsim\SI{500}{\tera\hertz}$ vs $\unsim\SI{5}{\giga\hertz}$; see Figure~\ref{fig:1_Bandwidth}a)\footnote{Small analog electronic circuits can have bandwidth $>\SI{5}{\giga\hertz}$ (e.g., Refs.~\cite{deal2016inp,thome2021first}) and small digital electronic circuits can be clocked at rates $>\SI{5}{\giga\hertz}$, but both analog and digital electronics for computing systems tend to be limited to speeds $\ll\SI{5}{\giga\hertz}$ by wire delays \cite{ho2001future,rabaey2002digital} and, since the mid-2000s, also by power dissipation \cite{horowitz2014computings}.}. This leads to two potential benefits:

\begin{enumerate}[label=(\roman*)]
    \item There is \textbf{massive frequency-multiplexing parallelism}, e.g., there can be $>10^7$ comb lines in a frequency comb \cite{diddams2020optical} and $>10^9$ frequency modes in a long fiber-ring cavity; data represented in each comb line (frequency mode) can be acted on in parallel (Figure~\ref{fig:1_Bandwidth}b)---not just individually (i.e., element-wise), but also with operations that, for example, add or multiply data in different frequency modes \cite{wright2022deep}. The parallelism of optical frequency modes is commonly taken advantage of in optical communications, where wavelength-division multiplexing enables communication over a single-mode fiber at rates $>10^{13}$ bits per second \cite{marin2017microresonator}; this technology can also be used for computing (e.g., Ref.~\cite{nakajima2021scalable}, which used a bandwidth of $B\sim\SI{5}{\tera\hertz}$).
    \item The \textbf{dynamics of optical systems can be very fast}, which can translate to very high operation speeds, which in turn can lead to higher computing throughput and lower latency\footnote{The Discussion section expands on what we mean by \textit{throughput} and \textit{latency} in computing.}: the limit in the delay for an operation, $\tau_\textrm{delay} \gtrsim 1/B$, can be $\unsim 100,000\times$ smaller for optics than electronics if the full bandwidth of optics is used.
    However, some subtlety is needed in the interpretation of this perspective on potential optical advantage from bandwidth. For one, the bandwidth limit on $\tau_\textrm{delay}$ is just a limit and the delay can be substantially longer than the limit if the device has a propagation length such that the time taken for light to travel through the device is long compared to $1/B$ (i.e., a speed-of-light limit begins to dominate; see also \circled{12}).\footnote{When the delay from propagation dominates the total delay, it is still possible to benefit from the fast bandwidth-limited speed in throughput by \textit{pipelining} \cite{hennessy2017computer}---e.g., by sending multiple optical pulses into the system spaced apart by more than the temporal pulse width $\unsim1/B$ but by less than the propagation delay. Since electronic computers can---and generally do---also take advantage of pipelining, care again needs to be taken in making performance comparisons.} For another, the delay for an individual modern electronic transistor under typical load is $\unsim\SI{1}{\pico\second}$ \cite{sicard2021introducing} so if one compared photonics to electronics at the level of an individual switch, the bandwidth benefit of optics would be much smaller than $\unsim 100,000\times$ (perhaps ``only'' $\unsim 1,000\times$). At the level of an entire chip, electronic processors are clocked $\unsim 10-100\times$ more slowly than the circuit delays \cite{xie2015performance} would suggest are possible, largely due to limits on power dissipation \cite{horowitz2014computings}. In contrast, photonic processors can have low dissipation (\circled{3}), and so at a system level it is a combination of intrinsic bandwidth \textit{and} low dissipation that gives rise to a $\unsim 100,000\times$ potential system-wide bandwidth advantage for optics.
    
    Optical switching of $\unsim\SI{46}{\femto\second}$ pulses has been demonstrated \cite{guo2022femtojoule}---highlighting the fast speeds possible with THz-bandwidth optical pulses and the quasi-instantaneous nature of nonlinear-optical operations.
\end{enumerate}

\end{enumerate}

\newpage

\begin{enumerate}[label=\protect\circled{\arabic*}]
\setcounter{enumi}{1}

\item \textbf{Spatial parallelism}: photonic systems can exploit a large number ($>10^6$) of parallel spatial modes \cite{kahn2017communications}. Consumer electronics using $>10^8$ spatial modes in a $\unsim \SI{2.5}{\centi\meter\squared}$ area have been realized \cite{samsung2022samsung}, illustrating that massive parallelism can be achieved in practice.

For photonic systems in which light is confined in a single two-dimensional plane, such as in two-dimensional photonic integrated circuits, the density of photonic \textit{components} can be as high as $\unsim 10^6$ per $\SI{}{\centi\meter\squared}$ \cite{bogaerts2020programmable}, and we can roughly think of each component as enabling $\geq 1$ computing operation (such as a multiplication) to be performed in parallel.\footnote{Why did we write $\geq 1$ operation and not just exactly $1$ operation? There are multiple reasons. For example, one is that a single component in space can act on many frequency modes in parallel, as mentioned in $\circled{1}$, or on multiple polarization modes. Another is that depending on one's definition of an operation, and one's definition of a single component, a component may naturally perform multiple operations in a single pass of light through it, such as a single 50:50 coupler arguably performing two multiplications and two additions.} While this component density is in absolute terms a high number, we should compare it against the spatial parallelism available in CMOS electronics, where the achieved density of transistors is $\unsim 10^{10}$ per $\SI{}{\centi\meter\squared}$ \cite{nvidia2022nvidia}.\footnote{As another point of comparison, to give an example of a candidate future electronics technology, an analog matrix-vector-multiplier core based on a crossbar array of phase-change memory, built by IBM \cite{khaddam2022hermes}, featured 65536 phase-change-memory cells within a chip area of $\unsim \SI{0.6}{\milli\meter\squared}$. This is a density of $\unsim 10^7$ cells per $\SI{}{\centi\meter\squared}$, and each cell can be interpreted as performing one scalar, analog multiplication per clock cycle.} In this setting of two-dimensional photonic integrated circuits, optics is at a disadvantage versus electronics in the pure density of fabricable components (since the transistor density in electronics is $\unsim 10^4 \times$ larger than the component density in on-chip photonics).\footnote{The density of transistors versus photonic components is arguably the most relevant comparison, since transistor-based electronic processors are, in most cases, the systems to beat. However, even two-dimensional photonics can have a spatial-parallelism advantage over two-dimensional microwave electronics: for example, photonic-crystal cavities (resonators) can have areas $\unsim\SI{1}{\micro\meter\squared}$ \cite{vahala2003optical,majumdar2012design}, whereas electronic microwave resonators are typically orders of magnitude larger (for example, see Ref.~\cite{da200610b}).} On the other hand, if the third spatial dimension is used \cite{psaltis1990holography,wetzstein2020inference}, optics may gain a several-orders-of-magnitude advantage\footnote{Let us use an example to make a rough estimate of the kind of advantage that is in principle possible. Consider a 2D photonic device with dimensions $L \times L$ and a 3D photonic device with dimensions $L \times L \times L$. Assume we address each device with light having wavelength $\lambda \approx \SI{500}{\nano\meter}$ and that the device length is $L \approx \SI{5}{\centi\meter}$. The number of resolvable spots in the former case is on the order of $(L / \lambda)^2 = 10^{10}$ whereas the number of resolvable voxels in the latter case is on the order of $(L / \lambda)^3 = 10^{15}$---an advantage of $(L / \lambda) = 10^5$ times when going from 2D to 3D. We can also compare these numbers with the counts of transistors in electronic processors: at the state-of-the-art fabrication density of $\unsim 10^{10}$ transistors per $\SI{}{\centi\meter\squared}$, a $\SI{5}{\centi\meter} \times \SI{5}{\centi\meter}$ chip would have $2.5 \times 10^{11}$ transistors. This is an order of magnitude greater than the number of resolvable spots in the same-area photonic device, but several orders of magnitude smaller than the number of voxels in the same-length 3D device. Of course an addressable voxel of material is not the same thing as a transistor; one ultimately needs to carefully analyze the computation and memory that is achieved using a particular device in a particular way, but these crude estimates hopefully convey two key messages: that by going from 2D to 3D devices, there can be an orders-of-magnitude increase in the achievable complexity of the device stemming from the fact that $(L / \lambda)$ can be a large number, and that while 2D photonic devices offer lower spatial parallelism than transistor-based electronic chips, moving to 3D devices may enable an orders-of-magnitude benefit in spatial parallelism for optics over electronics.} in spatial parallelism because electronics is in practice limited to very modest three-dimensional integration.\footnote{A typical modern electronic chip is thin---on the order of $\SI{1}{\milli\meter}$---and comprises only tens of layers \cite{sell2022intel}, whereas optical processors that are centimeters or even meters thick, e.g., using propagation through bulk crystals \cite{psaltis1990holography,li1993optical} or multimode optical fiber \cite{teugin2021scalable}, have been constructed. In the specific case of NAND memory, electronic integrated circuits have been scaled to 128 layers \cite{goda2020_3d}---which suggests that for memory rather than computing, photonics has less room for advantage over electronics by extending in the third dimension.}

However, there is an important additional perspective on spatial parallelism: it is not only the density or number of components you can fabricate that is important, but how many of the components you can in practice use in parallel. In other words, increased component density does not necessarily translate to proportionately greater computing performance. Modern CMOS electronic processors are typically only able to switch a small percentage (e.g., 3\% \cite{dally2022future}) of their transistors in a single clock cycle, largely due to limitations in cooling \cite{hennessy2017computer}. When taking into account how many components can actually be operated in parallel with the constraints of power dissipation (see also \circled{3}), two-dimensional photonic integrated circuits may not be at as much of a disadvantage in spatial parallelism versus electronic integrated circuits as the fabricated component densities alone would suggest.

As an example of spatial parallelism in optical computing, free-space optical processors have been prototyped using commercial spatial light modulators, which have $\unsim 10^6$--$10^7$ controllable pixels---making them useful tools in building highly parallel systems \cite{neff1990two}. Refs.~\cite{zhou2021large,wang2022optical} both demonstrate the use of $\unsim 5\times 10^5$ pixels to compute $\unsim 5\times 10^5$ scalar multiplications in parallel per pass of light through their optical setup. For applications in which the programmability of spatial light modulators is not required (such as in neural-network inference), fabricated metasurfaces offer a route to even larger parallelism: based on the linear-with-area scaling of the space--bandwidth product of imaging systems \cite{mait2018computational}, we expect it to be possible to perform metasurface-based matrix multiplications or convolutions with $>10^9$ preprogrammed pixels (parameters) on $\unsim \SI{100}{\centi\meter\squared}$ surfaces \cite{colburn2019optical}.

\end{enumerate}

\begin{figure}[h]
  \includegraphics[width=0.5\textwidth]{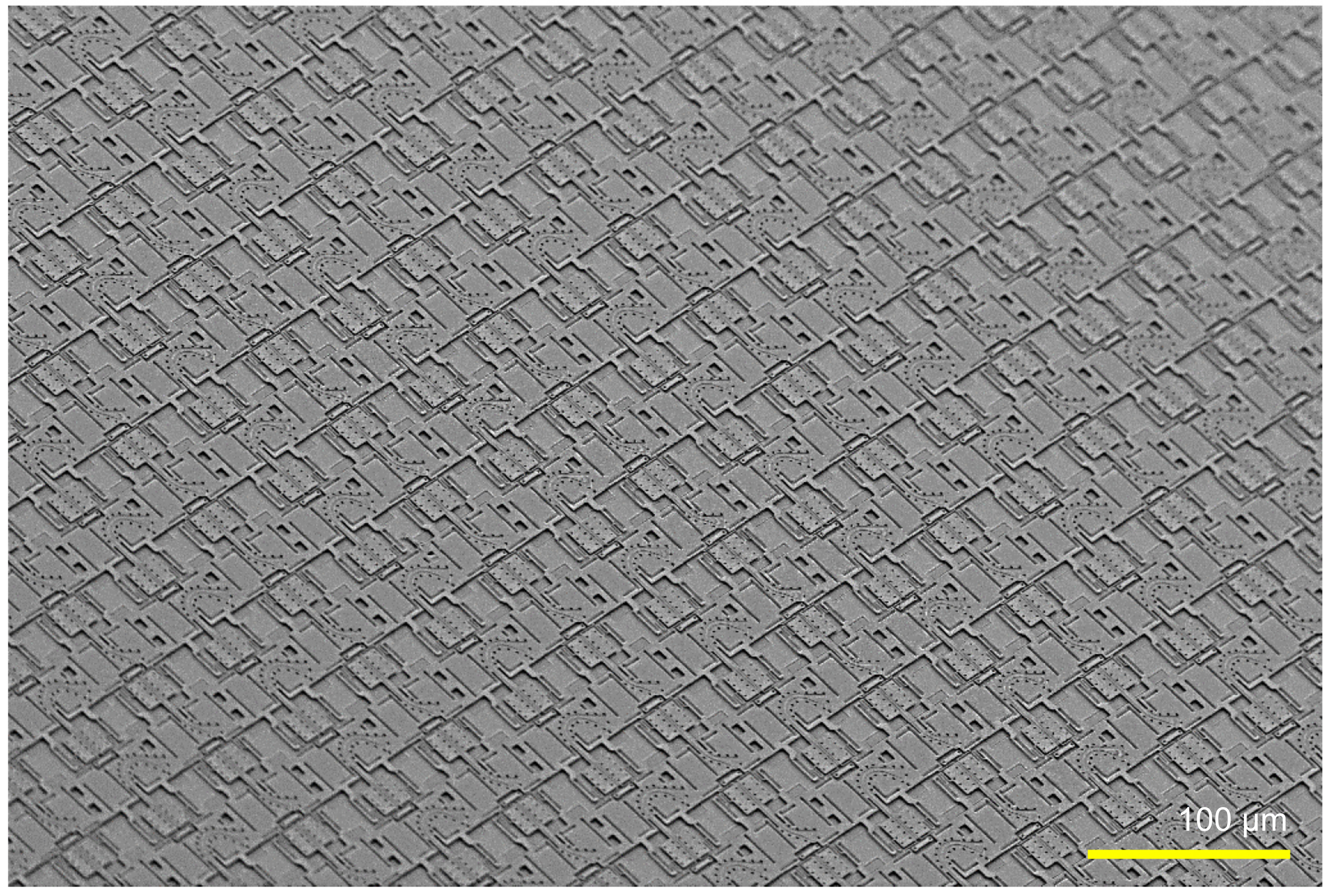}
  \caption{\label{fig:2_SpatialParallelism} \protect\circled{2} \textbf{Spatial parallelism.} Part of a state-of-the-art silicon-photonic device with 16,384 pixels on a $10 \times 11\textrm{-mm}^2$ chip, illustrating the degree of spatial parallelism possible in modern photonic devices. {\color{gray} Adapted from Ref.~\cite{zhang2022large}, \textcopyright~Springer Nature.}}
\end{figure}

\newpage

\begin{enumerate}[label=\protect\circled{\arabic*}]
\setcounter{enumi}{2}

\item \textbf{Nearly dissipationless dynamics}: photons can propagate through free-space (or even some on-chip\footnote{For example, thin-film lithium niobate chips can have waveguide propagation losses of $\SI{0.06}{\decibel\per\centi\meter}$ \cite{desiatov2019ultra}.}) optical setups with nearly no energy loss, and perform computation by their mere propagation. How much computation? We consider the cases of \textit{linear-} and \textit{nonlinear-optical} systems:

\textit{Linear optics}: an example of this phenomenon is that a single lens effectively performs a two-dimensional Fourier transform on light that impinges on it \cite{goodman2004introduction}---optical correlators \cite{ambs2010optical} and convolutional layers in optical neural networks \cite{wetzstein2020inference} both take advantage of this. More generally, propagation of light through a linear-optical system can be modeled by a matrix-vector multiplication, so matrix-vector multiplication can be performed by merely shining light encoding a vector (of dimension $N$) in its spatial\footnote{For the sake of concreteness, in this paragraph we give examples of vectors encoded in space, but this is not the only possibility: the propagation of light in just a single spatial mode can also result in nearly dissipationless computation of inputs encoded in other ways, such as in frequency or time \cite{shastri2021photonics}.} pattern onto an optical system \cite{wetzstein2020inference}. As a rather extreme example, shining light through white paint can be used to perform the multiplication of a vector by a random matrix with dimension $>10^6 \times 10^6$ \cite{saade2016random}. A variety of linear-optical systems in which the matrix can be programmed\footnote{As opposed to the example of the multiplication using light propagation through white paint, in which the matrix is fixed and random.} have also been demonstrated \cite{bogaerts2020programmable,wetzstein2020inference}, although in these cases the matrix size has generally been limited by the number of programmable elements (as such spatial-light-modulator pixels\footnote{Spatial light modulators with $\unsim 10^7$ pixels are commercially available; each pixel can be used to represent a single programmable element of a matrix.}) that can be engineered. In principle, the dissipationless nature of optical propagation can lead to matrix-vector multiplications being performed that beat the Landauer limit \cite{lent2018energy} for multiplications performed on digital electronic processors---intuitively because in a coherent setup, the optical interference that occurs is a reversible process \cite{hamerly2019large}.

\textit{Nonlinear optics}: propagation of light through nonlinear-optical systems can also exhibit nearly dissipationless dynamics that can be harnessed for computation. For example, propagation of light through an optical medium with a nonzero second-order nonlinear-optical susceptibility, $\chi^{(2)}$, can in general result in sum-frequency- and difference-frequency-generation processes where the optical amplitude of the output scales as the product of the amplitudes of light at two frequencies at the input, e.g., $E_\textrm{out}(\omega_1 + \omega_2) \propto  E_\textrm{in}(\omega_1) E_\textrm{in}(\omega_2)$ \cite{boyd2020nonlinear}. We can interpret such a nonlinear-optical process as performing a scalar multiplication of the two numbers $E_\textrm{in}(\omega_1)$ and $E_\textrm{in}(\omega_2)$ \cite{wright2022deep}. Nonlinear-optical dynamics enable the implementation of mathematical functions that are nonlinear---which is essential in deep neural networks \cite{goodfellow2016deep} and in computing more generally \cite{kia2017nonlinear}. For example, in a $\chi^{(2)}$ process, if the frequencies of the input light are equal ($\omega_1 = \omega_2$), then one may obtain output light at twice the frequency with amplitude $E_\textrm{out}(2 \omega_1) \propto  (E_\textrm{in}(\omega_1))^2$, so the function realized is $f(x)=x^2$, which is nonlinear. Furthermore, just as the propagation of multiple spatial beams through a linear-optical system can be seen as performing a matrix-vector product, propagation of multiple spatial beams through a nonlinear-optical system can realize a higher-dimensional generalization of matrix-vector multiplication, namely tensor contraction involving tensors of order $n+1$, where $n$ is the order of the nonlinearity-optical susceptibility, $\chi^{(n)}$. This is an impressive feature for computing \cite{teugin2021scalable,wright2022deep}: with the lowest-order nonlinearity, $n=2$, the computation performed---again, by the mere propagation of the light through the system---is a tensor contraction that comprises $\unsim N^3$ multiplication operations, where $N$ is again the number of spatial modes. Higher orders of optical nonlinearity can result in even larger amounts of computation being performed by a single pass of light through the system, since even-higher-order tensors are involved.

The fact that computations can be performed nearly dissipationlessly in optics has two potential benefits:

\begin{enumerate}[label=(\roman*)]
\item \textbf{Higher energy efficiency}: the obvious benefit is that one can potentially harness dissipationless dynamics to perform computation using less energy than would have been needed in a different platform that did have substantial dissipation (such as electronics).

\item \textbf{Higher performance}: dissipation doesn't just cause a computation to cost more energy, but can also limit the clock speed and parallelism of a processor, ultimately limiting its total computing throughput (operations per second) and latency. Modern CMOS electronics processors are limited---both in clock speed and in three-dimensional density of transistors---by our ability to extract dissipated heat from them \cite{horowitz2014computings}.\footnote{Photonics has another potential benefit over electronics with regards to extracting heat from dissipation within a three-dimensional chip: whereas the loss of electrical energy in a chip is generally by the generation of heat at the point where the energy is lost---e.g., resistive heating of a wire---the situation in photonics can be quite different because the loss of optical energy is often \textit{not} due to absorption and accompanying generation of heat, but rather by scattering. This is true for waveguides in silicon photonics integrated circuits, for example, and suggests that if you construct a three-dimensional silicon-photonic chip, the losses of waveguides within the chip will primarily not cause heating, but instead will result in photons being scattered within the chip until they emerge at the surfaces. In summary, nearly dissipationless dynamics in optics enables us to create three-dimensional photonic chips that don't suffer from the extreme heat-extraction challenges of three-dimensional electronic chips, and even the small photonic dissipation that does occur does not cause heating within the bulk of the chip if it is due to scattering, so we may not even need to worry about the residual photon loss causing heat-management difficulties provided that components that absorb photons are avoided.} By dramatically reducing dissipation per computing operation, one potentially allows for a dramatic increase in both the clock speed and spatial parallelism (number of operations performed simultaneously per unit volume).

\end{enumerate}

There is however a snag, namely \textit{input/output costs}: how does the input data for the computation get loaded and the result get read out? If the input comes from an electronic memory and the result needs to be stored in an electronic memory then even though the computation itself can happen nearly ``for free'', one needs to convert electronic data to the optical domain for the data input, and then convert the optical answer back to the electronic domain. This memory access and transduction, which will typically also involve digital-to-analog and analog-to-digital conversion, will cost substantial energy (and be limited in speed when compared to optical bandwidths of THz). Fortunately this energy cost will only scale as the size of the input vector, $N$, whereas the amount of computation being performed may scale as $N^2$ (linear propagation) or $N^3$ (or even higher powers; nonlinear propagation), and so for sufficiently large $N$, the energy cost of the input/output will be small compared to the cost that the computation would have required in an electronic processor. Similarly, the time required for input/output for $N$-dimensional vectors can, for sufficiently large $N$, be very small compared to the time the $N^2$- or $N^3$-complexity computation would have taken on an electronic processor. The loading of coefficients, such as the matrix elements in the case of linear propagation, in general also has a cost in both energy and time but this can be amortized over many runs, such as in the case of batched inference with neural networks \cite{shen2017deep}.

\end{enumerate}

\begin{figure}[h]
  \includegraphics[width=0.5\textwidth]{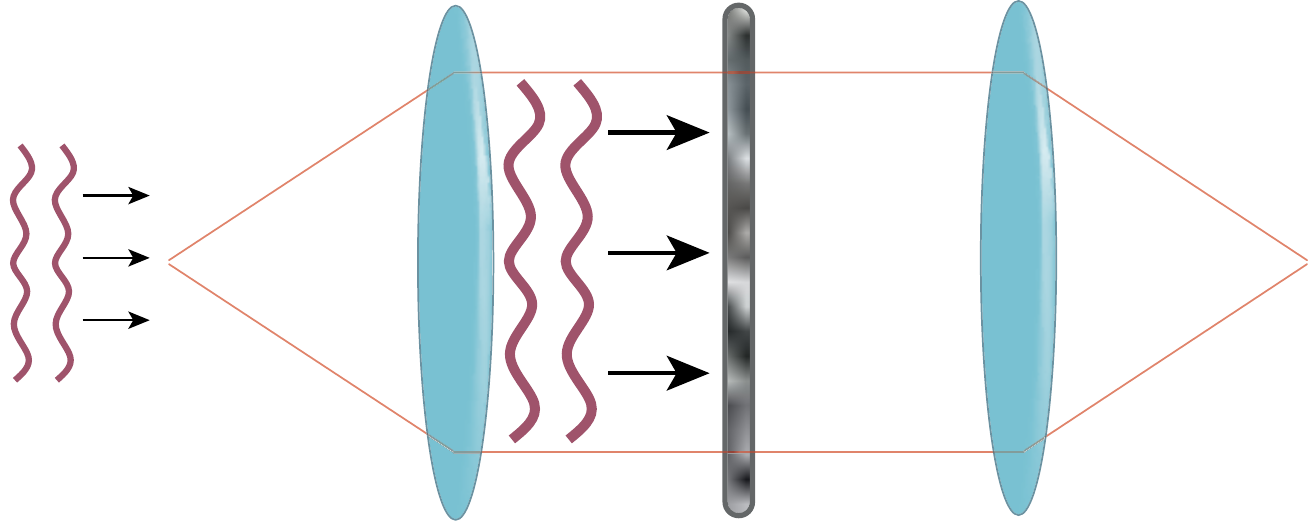}
  \caption{\label{fig:3_DissipationlessDynamics} \protect\circled{3} \textbf{Nearly dissipationless dynamics.} An example of computing with \textit{linear} optics: light propagating through a lens undergoes a Fourier transform, and in a two-lens $4f$ system with a scattering medium inbetween, a convolution is performed on the input light. In the absence of optical loss (e.g., from absorption in the lenses), the computation of the convolution happens without any energy loss. However, if one considers how to use this building block in an end-to-end computing system, there \textit{is} typically an energy cost associated with converting an electrical signal into an optical input, and there is also typically an energy cost associated with converting the optical output back into an electrical signal. {\color{gray} Adapted from Ref.~\cite{wetzstein2020inference}, \textcopyright~Springer Nature.}}
\end{figure}

\newpage

\begin{figure}[h]
  \includegraphics[width=0.48\textwidth]{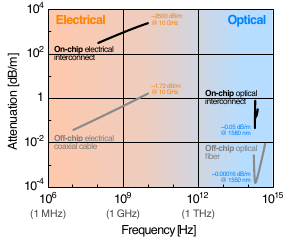}
  \caption{\label{fig:4_LowLossTransmission} \protect\circled{4} \textbf{Low-loss transmission.} For both on-chip and off-chip transmission, the signal attenuation (in dB per meter) is orders of magnitude lower (better) with optical instead of electrical signals. For example, electrical signals at $\SI{10}{\giga\hertz}$ have $\unsim 10^4\times$ higher attenuation than equivalent on-chip or off-chip transmission with optical signals. {\color{gray} Inspired by Ref.~\cite[Figure~4.3]{stallings2007data}. \textit{Data sources}: On-chip electrical interconnect: Ref.~\cite{kleveland2002high}; Off-chip electrical coaxial cable: Ref.~\cite{qaxialRG142B}; On-chip optical interconnect: Ref.~\cite{bauters2011planar}; Off-chip optical fiber: Ref.~\cite[Figure~22.2]{schubert2012light} and Ref.~\cite{corningSMF28}. This figure is intended to give a heuristic comparison; it does not comprehensively cover all transmission technologies, but is based on just a few illustrative examples that convey the relevant orders of magnitude. For more examples and details, see: Ref.~\cite{miller1997limit} (electrical interconnects and cables); Ref.~\cite{kleveland2002high} (on-chip electrical interconnects with different dimensions); Ref.~\cite{huang2003optical} (electrical interconnects on printed circuit boards); Ref.~\cite{shams2022reduced} (integrated-photonics waveguides with lithium niobate).}}
\end{figure}

\begin{enumerate}[label=\protect\circled{\arabic*}]
\setcounter{enumi}{3}

\item \textbf{Low-loss transmission}: the energy cost to transmit information ``long'' distances with light is much lower than with electrical signals \cite{miller2017attojoule}, mostly\footnote{There are several subtleties in evaluating the energy cost of optical and electrical communication, discussed in detail in Refs.~\cite{ho2001future,jose2006pulsed,miller2017attojoule}, which necessitate the use of the word ``mostly'' here. For one, optical communications between electronic devices require transduction of signals from electrical to optical, and back to electrical, and the transduction devices have energy costs \cite{miller2017attojoule}. For another, electrical signal transmission along a wire requires energy that increases with length because the wire's resistance increases with length---but this is not the end of the story: for thin wires, such as those used in CMOS electronic processors, the wire delay grows quadratically with length and to mitigate this, repeaters are used to regain a linear scaling of delay with length, and the repeaters also have an energy cost (associated with the switching of their driver transistors) \cite{ho2001future,jose2006pulsed}.} because signal attenuation (energy loss) per unit length is much higher in electrical wires than in optical fibers or waveguides (Figure~\ref{fig:4_LowLossTransmission}).

For on-chip photonic processors, commercial foundries such as AIM Photonics can produce silicon-nitride waveguides with losses $\sim\SI{0.06}{\dB\per\centi\meter}$ for wavelengths $\sim\SIrange{1600}{1640}{\nano\meter}$ and $<\SI{0.25}{\dB\per\centi\meter}$ across the telecommunications C band ($\sim\SIrange{1530}{1565}{\nano\meter}$) \cite{tyndall2022low}.

An important caveat for both free-space and on-chip optical processors is that while propagation losses between components can be very low, typically there will be losses from reflections or scattering as light propagates into or out of a component (e.g., Fresnel reflections due to mismatch in refractive index). As a result, optical processors still need careful design to avoid excessive overall optical loss.

The low-loss transmission of optics is already being taken advantage of in electronic computing: optical links in datacenters \cite{cheng2020optical}, and even directly between chips \cite{sun2015single}, use light to communicate information over length scales from centimeters to many meters. It is anticipated that even some communications within a single chip might eventually use optics \cite{miller2017attojoule,cheng2020optical}.

A major reason that light is not already used for communications within single electronic-processor chips, especially over very short distances, is that the optoelectronic components to transduce signals between the optical and electrical domains cost both space and energy, and it is only worth paying these costs when the distance the signal needs to travel is long enough \cite{miller2017attojoule}. An optical computer, on the other hand, could in principle take advantage of optics for low-energy-cost, nearly-dissipationless information transmission at \textit{all} length scales, and without paying space or energy costs for transduction\footnote{An optical processor will inevitably need to use some energy for transduction, for example to load the initial input data for the computation and/or to read out the final answer, which will typically need to be in the electrical domain. But the transductions---and their costs---that would have occurred within a computation can be avoided.}---because the signals would already be optical.

\end{enumerate}

\newpage

\begin{figure}[h]
  \includegraphics[width=0.25\textwidth]{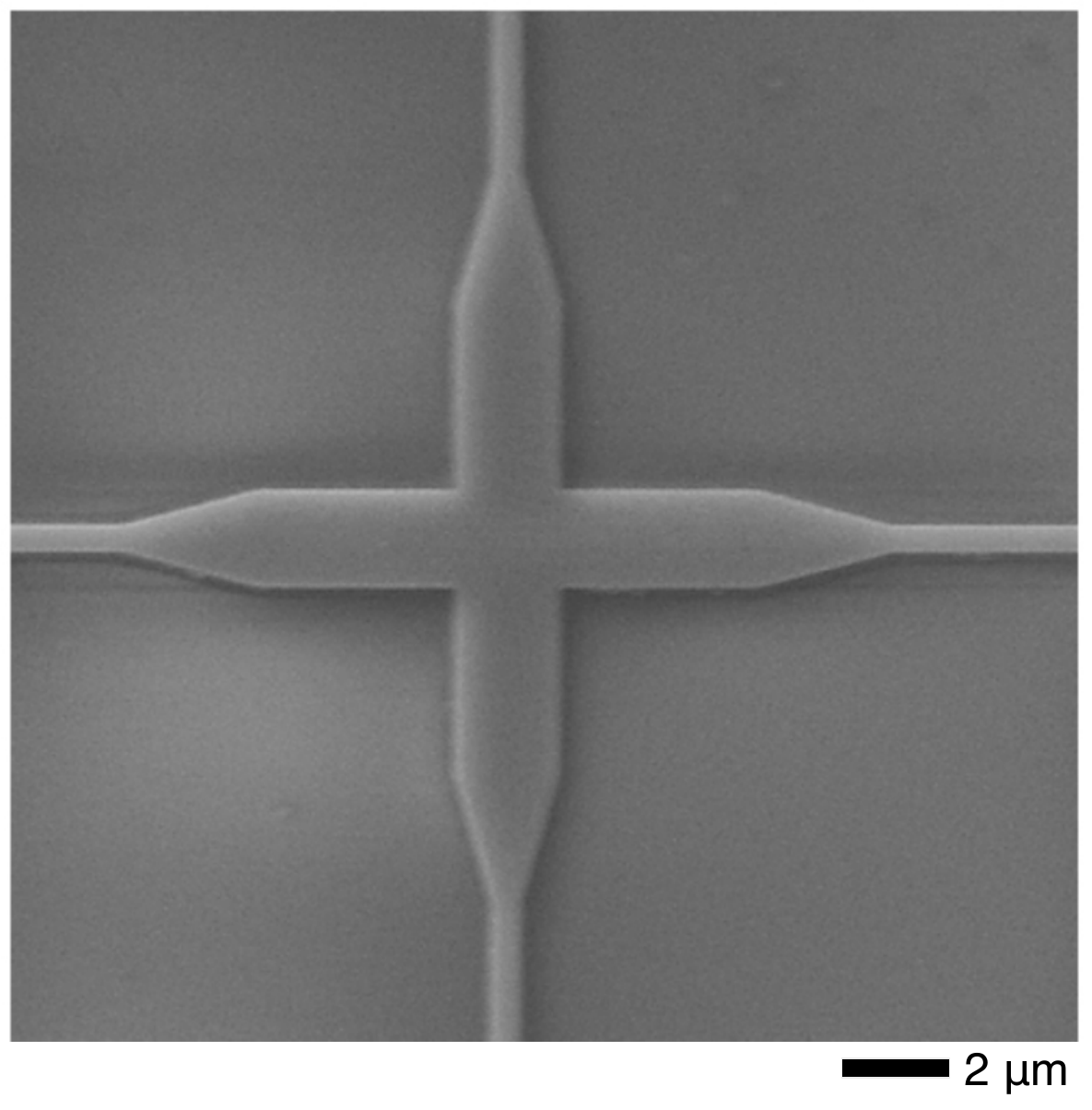}
  \caption{\label{fig:5_OpticalBeamsCanCross} \protect\circled{5} \textbf{Optical beams and ``wires'' can cross.} It is not only in free space that optical paths can cross: in integrated photonics, waveguides can pass through one another with minimal impact on the signal propagation. The waveguide crossing in this figure had a crosstalk of $<\SI{-50}{\dB}$. {\color{gray} Adapted from Ref.~\cite{johnson2020low}, \textcopyright~Optica Publishing.}}
\end{figure}

\begin{figure}[h]
  \includegraphics[width=0.5\textwidth]{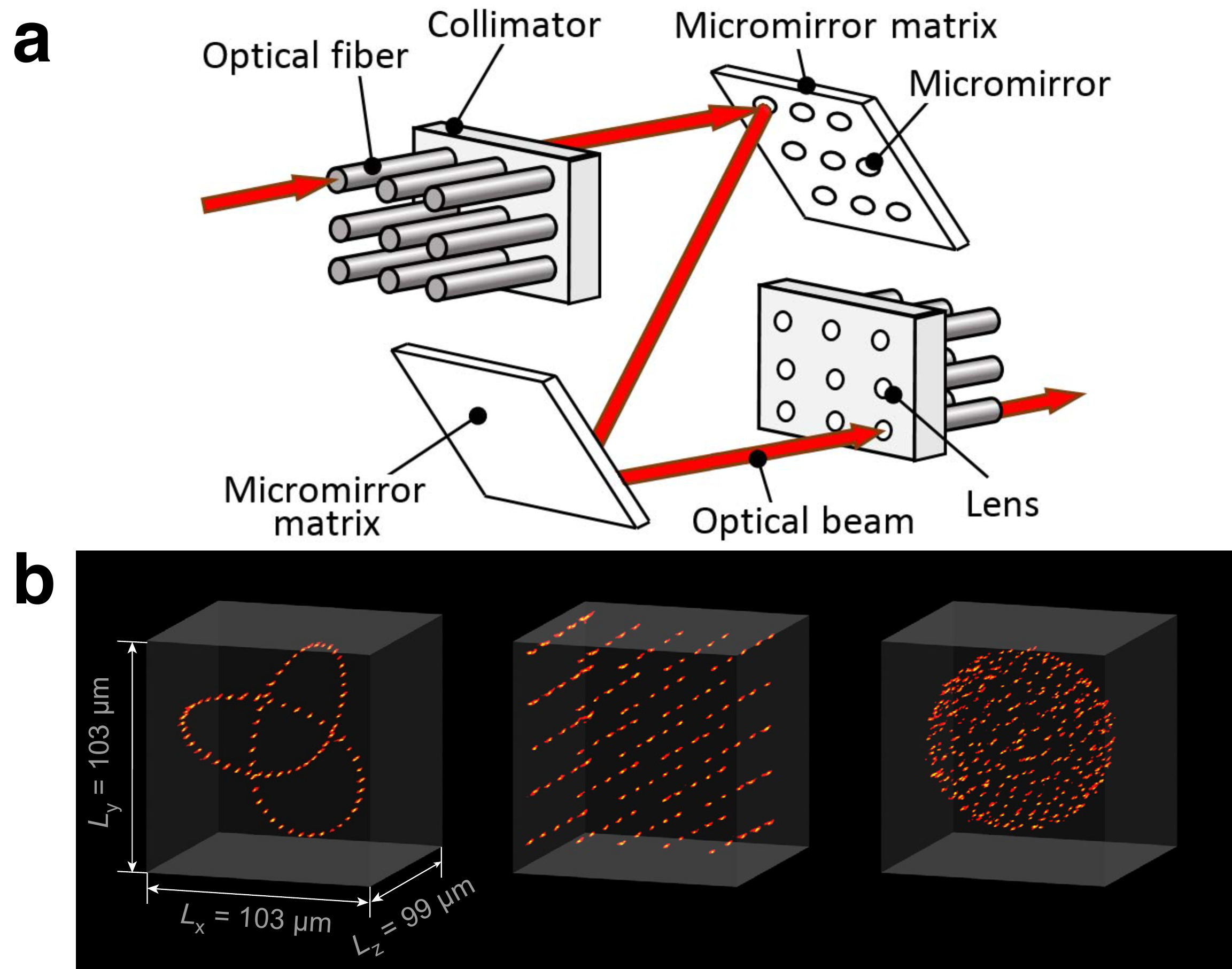}
  \caption{\label{fig:6_OpticalBeamsSteered} \protect\circled{6} \textbf{Optical beams can be steered programmably.} \textbf{a}, Optical beams inside a micro-electro-mechanical systems (MEMS) optical switch can be rerouted on timescales on the order of milliseconds using arrays of MEMS-actuated micromirrors. \textbf{b}, Optical-tweezer beams can be reconfigured to trap atoms in arbitrary geometries in 3D; the results shown here are from an experiment in which a liquid-crystal-based spatial light modulator was used to program the beams; such modulators can also updated on a timescale on the order of milliseconds. {\color{gray} Part \textbf{a} reproduced from Ref.~\cite{stepanovsky2019comparative}, \textcopyright~IEEE. Part \textbf{b} adapted from Ref.~\cite{barredo2018synthetic}, \textcopyright~Springer Nature.}}
\end{figure}

\begin{figure}[h]
  \includegraphics[width=0.5\textwidth]{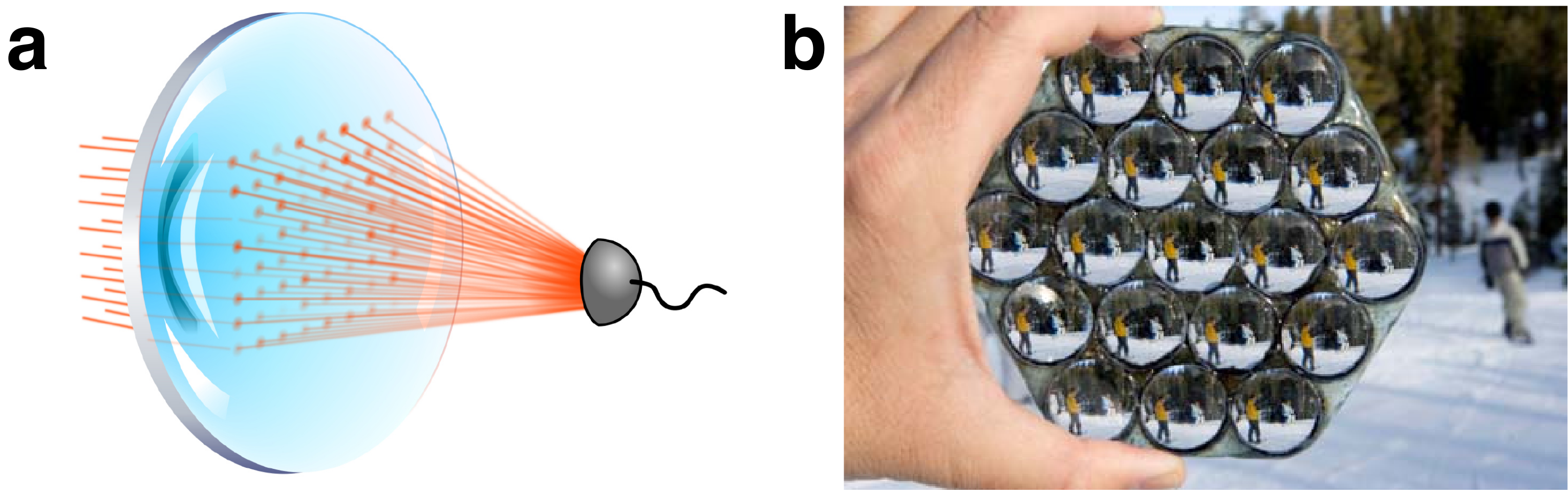}
  \caption{\label{fig:7_FanInFanOut} \protect\circled{7} \textbf{Optical fan-in and fan-out.} \textbf{a}, Fan-in can be performed in free space using a lens; here, a lens causes many beams to converge on a single-pixel detector. \textbf{b}, Fan-out can be performed in free space using an array of lenses, where each lens ``captures'' a copy of the incoming image. {\color{gray} Part \textbf{a} adapted from Ref.~\cite{wang2022optical}, \textcopyright~Springer Nature. Part \textbf{b} reproduced from Ref.~\cite{georgiev2006light}, \textcopyright~Adobe Systems.}}
\end{figure}

\begin{enumerate}[label=\protect\circled{\arabic*}]
\setcounter{enumi}{4}

\item \textbf{Optical beams and ``wires'' can cross whereas electrical wires cannot}: optical beams can pass through one another without suffering from cross-talk\footnote{This is true under the condition of negligible optical nonlinearity---which is often the case, especially in free-space settings, but also in materials when the optical power is low and the propagation length is short. In other words, informally: we don't have lightsabers in ordinary optical situations \cite{fillion2019physical}.}, and optical on-chip ``wires'' (waveguides; see Figure~\ref{fig:5_OpticalBeamsCanCross}) can cross with very low cross-talk---not just in principle, but also in practice in the presence of fabrication imperfections. In contrast, electrical wires need their own region of isolated physical space, and in addition to not being able to pass through one another, also often suffer from cross-talk even if they are merely close to one another \cite{duan2010and}. This provides the possibility for photonic processors to be more compact than electronic processors when interconnect is an important contributor to processor size, although the use of optical beams for communicating information is not without its own cross-talk challenges due to diffraction, scattering, and unwanted reflections \cite{lee1995design}.\footnote{One might also wonder about the size of optical beams versus electrical wires, since optical beams or waveguides will be limited to sizes on the order of a wavelength $\lambda$, whereas electrical wires can be made only nanometers wide. However, interconnects in electronic processors have trace (``wire'') widths and spacings on the order of $\SI{1}{\micro\meter}$ \cite{nassif2022sapphire}, which is a design choice in part motivated by the fact that a wire's resistance decreases as its cross-sectional area increases \cite{ho2001future}.}

One can interpret the ability for optical beams to cross as a key enabler of many free-space, spatially-multiplexed optical implementations of convolution and matrix-vector multiplication \cite{wetzstein2020inference}. For example, in implementations (e.g., Ref.~\cite{wang2023image}) of matrix-vector multipliers that use arrays of lenses for fan-out (Figure~\ref{fig:7_FanInFanOut}b), the rays between the input vector and the fanned-out copies cross. The crossing supports the implementation, in principle, of large convolutions and dense matrix-vector multiplications in small volumes. Optical switches such as the one shown in Figure~\ref{fig:6_OpticalBeamsSteered}a provide another example of where crossing of beams enables a more compact design.

\end{enumerate}

\newpage

\begin{enumerate}[label=\protect\circled{\arabic*}]
\setcounter{enumi}{5}

\item \textbf{Optical beams can be steered programmably at high speed whereas electrical wires are either fixed or reconfigureable only slowly}: free-space optical beams can easily be redirected (for example, using an acousto-optic deflector, with a delay on the order of microseconds), enabling the creation of reconfigurable optical interconnects \cite{goodman1985fan,mcardle2000reconfigurable}. In contrast, electrical wires on chips are fixed at the time of fabrication, and wires joining nodes in an interconnect between processors, boards, or racks can only be moved slowly (typically on the order of seconds).\footnote{How do electronic processors deal with quasi-fixed interconnects? The disadvantage of having a fixed network is typically mitigated by using multi-hop communications---relying on there being a path between a sender and a receiver involving some intermediate nodes---and switching, which achieves fast rerouting of signals within a fixed network topology. These strategies come with the cost of increased latency and potential bandwidth bottlenecks.}

\end{enumerate}

\newpage

\begin{enumerate}[label=\protect\circled{\arabic*}]
\setcounter{enumi}{6}

\item \textbf{Fan-in (summation) and fan-out (copying) work differently in optics}: copying data to be processed in parallel (fan-out), and summing the outputs from a number of parallel-processing units (fan-in) are important primitives in parallel processing. Both can be implemented in optics in a way that is different than in electronics, and have different tradeoffs \cite{goodman1985fan,wang1990limits}. Optics has a potential advantage from supporting large ($>1000$) fan-in and fan-out without the $RC$ and $LC$ delays\footnote{As Ref.~\cite{goodman1985fan} points out, when evaluating an optical scheme, one needs to take care to evaluate the $RC$ and $LC$ delays of photodetectors that are involved.} of fan-in/fan-out with electrical wires, for which fan-in/fan-out is typically kept lower than $10$ in digital processors, necessitating multiple buffering stages (and hence further delay) whenever larger fan-in/fan-out is needed \cite{ji1997impact,chen2006maximum}.

In free space, fan-in of signals encoded in spatial modes can be performed by directing beams to a common point in space (e.g., via the use of a lens; see Figure~\ref{fig:7_FanInFanOut}a), at which there could be, for example, a photodetector (if the next processing step required conversion from optical to electrical signals), a holographic element (to combine the beams traveling in different directions into a beam that travels in one direction, albeit at the cost of loss of optical power) \cite{goodman1985fan}, or an intensifier (which can amplify the summed beams and re-emit a single optical signal) \cite{wang2023image}.

Fan-out of a signal in a single spatial mode to multiple spatial modes can also be performed conceptually easily in free space, where it happens essentially without any special engineering effort (Figure~\ref{fig:7_FanInFanOut}b): imagine an optical display (such as a light-emitting-diode display on a cell phone) that emits in multiple directions---multiple people looking at the display from different vantage points can all see the same image, and we can interpret what happened is that multiple copies of the data on the display were made and transmitted to different receivers (people).\footnote{Another example of optical fan-out in everyday life is in a kaleidoscope.} Arrays of lenslets (microlenses) can be used to collimate the image copies \cite{de1989adaptive,wang2023image}.\footnote{Free-space fan-out can also be implemented and understood in the Fourier domain \cite{bernstein2022single}.}

Both fan-in and fan-out for spatial modes can also readily be implemented in integrated-photonics platforms \cite{yao2021compact}. However, in an on-chip setting light propagation is typically practically restricted to be in a single plane, whereas in free space it is natural for signals to propagate in all three dimensions, enabling a much higher degree of fan-in and fan-out. For this reason, it is easier to imagine gaining an advantage over on-chip electronic processors (which are also quasi-planar) from the use of optical fan-in or fan-out in free-space settings.

Up to here, we have discussed fan-in and fan-out in the context of spatial modes. For optical computers using frequency or temporal modes, fan-in and fan-out may be realized using other means than the spatial approaches referred to so far. For example, fan-out of data input as electronic signals can be performed in the frequency domain by modulating an optical frequency comb \cite{xu2021tops}, and weighted fan-in can be performed using wavelength-division multiplexing, including in on-chip platforms \cite{shastri2021photonics}.

To reason about \textit{why} or \textit{when} optical fan-in or fan-out may have an advantage over electrical fan-in or fan-out, it is useful to consider the bandwidth (\circled{1}) and low-loss transmission possible in optics (\circled{4}), and that optical beams can cross (\circled{5}).\footnote{Teasing out the source of a potential advantage can be quite subtle. For example, fan-in arguably plays an important role in enabling vector-vector or matrix-vector multiplication engines that use extremely small amounts of optical energy per multiplication \cite{wang2022optical}---in which the amount of optical energy needed to achieve a particular signal-to-noise ratio for a vector-vector dot product is fixed regardless of the vector size---but similar efficiency can be achieved with optoelectronic fan-in, in which summation is performed in the electrical domain \cite{hamerly2019large,sludds2022delocalized}. Purely analog electronic approaches to compute vector-vector dot products can also show favorable energy consumption versus digital electronic approaches \cite{murmann2015mixed}, so for any given computing scheme using optical fan-in, one can ask: which part of the potential benefit comes from performing the summation in an analog rather than digital fashion, and which part comes from using optics instead of electronics?} However, the fan-in/fan-out possibilities of optics (this point, \circled{6}) are distinct from the potential benefits of bandwidth, low-loss transmission and beam-crossing in optics, and it is fruitful to think of fan-in and fan-out in optics as special features that can be used in an optical-computing architecture, even though they may also use other features of optics to operate well.

\end{enumerate}

\newpage

\begin{figure}[h]
  \includegraphics[width=0.4\textwidth]{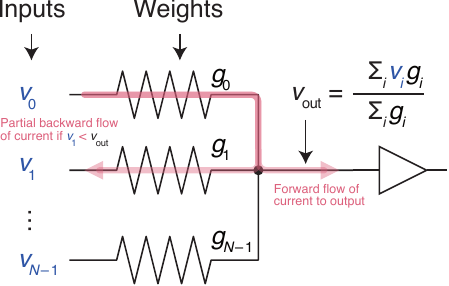}
  \caption{\label{fig:8_OneWay} \protect\circled{8} \textbf{One-way propagation.} An electrical fan-in (weighted sum of voltage inputs $v_i$ by conductance weights $g_i$) exhibiting undesired backwards flow of current. The current contributions from the input $v_0$ to the output (desired) and, if $v_1 < v_\textrm{out}$, from the input $v_0$ to the input $v_1$ (undesired) are shown in pink. Only the current contributions from $v_0$ to the output and to $v_1$ are illustrated here, but in general current will flow backwards from the common node $v_\textrm{out}$ to $v_i$ if $v_i < v_\textrm{out}$. In contrast, one-way, forward-only propagation of light in a fan-in is shown in Figure~\ref{fig:7_FanInFanOut}(a). {\color{gray} Adapted from Ref.~\cite{debenedictis2016computational}, \textcopyright~IEEE.}}
\end{figure}

\begin{enumerate}[label=\protect\circled{\arabic*}]
\setcounter{enumi}{7}

\item \textbf{One-way propagation}: light naturally propagates in one direction, from its source to a receiver\footnote{More pedantically, one can construct optical systems in which the propagation is naturally one-way; if one, for example, forms an optical cavity in part of the system, then the situation becomes more complicated.}, whereas electrical signals can propagate backwards (Figure~\ref{fig:8_OneWay}). In electronic processors, backwards propagation (from inputs to other inputs, or from the output to the inputs) can cause unwanted dynamics as well as unnecessary power consumption. This leads to an advantage of optics over electronics for some analog architectures.

While backwards propagation is a general feature of electrical circuits---without isolating elements such as buffers or diodes in a circuit, any time there is a voltage difference between two connected circuit nodes there will be a current flow between them, even if those two nodes are inputs---concerns about backwards propagation have arisen mostly in the context of analog crossbar-array processors, related to their fan-in stage \cite{debenedictis2016computational} and also the \textit{sneak path} issue \cite{shi2020research}. Analog \textit{optical} matrix-vector-product engines \cite{wetzstein2020inference} generally feature one-way propagation, avoiding some of the issues that arise in analog \textit{electronic} matrix-vector-product engines (i.e., crossbar arrays), and there is a broader notion of optics providing natural isolation \cite{aluf2012optoisolation} that can be useful in computing.

A caveat is that while perfectly one-way propagation is possible if light does not pass through any interfaces, any useful optical processor will involve at least some interfaces (e.g., light going from air into a glass lens), and as a consequence have some unavoidable reflections. The reflections can be made small by appropriate choices of geometry and materials but will never be completely eliminated. In many cases there may be an engineering tradeoff between, for example, the compactness of the optical processor and the magnitude of the reflections (i.e., the one-way-ness) in the system.

\end{enumerate}

\newpage

\begin{enumerate}[label=\protect\circled{\arabic*}]
\setcounter{enumi}{8}

\item \textbf{Adiabatic, least-action and least-power-dissipation principles in physics have different realizations in optical and electrical systems}: there are general physics principles---such as adiabaticity, the Principle of Least Action and the Principle of Least Energy Dissipation---that can lead to a physical system heuristically solving optimization problems \cite{vadlamani2020physics}, and different variations of these principles can be leveraged to construct optimization machines (such as Ising machines \cite{mohseni2022ising})\footnote{Given how central optimization is in machine learning, and especially in neural networks, computers designed to perform optimization are often also well-suited to perform machine learning---so an advantage on optimization can quite plausibly be translated into an advantage in machine learning too. Similarly, one can recast the problem of solving partial differential equations as a variational optimization problem \cite{e2018deep}, providing another potential application of physics optimization principles to a broader class of computations.}.

For example, Fermat's Principle of Least Time for optics states that light follows the path that minimizes its time to travel between two points\footnote{Feynman gave an explanation of this principle with a path-integral formulation in which the light can take all possible paths but only the paths that constructively interfere contribute substantially, and paths with substantially different propagation times than Fermat's solution destructively interfere \cite{feynman2006qed}. This perspective is possibly helpful for thinking about how to design optimization machines that use Fermat's principle.}, but this principle doesn't have a direct analog in electrical circuits---so a computer performing optimization using Fermat's principle is more natural to try create with optics.

Onsager's Principle of Least Energy Dissipation can apply in both optics and in electronics, but the behavior and resulting computing performance may be different because of differences in the underlying physics. For example, lasers and parametric oscillators in optics have a threshold when gain is equal to loss, and the fact that they will first oscillate in the mode with lowest loss can be used to design optical Ising machines \cite{wen2012injection,marandi2014network}. Electrical circuits, including oscillators, also have dynamics that heuristically minimize the energy dissipated \cite{vadlamani2020physics}, but they are not identical to lasers or optical parametric oscillators and in general will have different behavior. It is an open question whether, or in which situations, optics systems using Onsager's principle have an advantage over electronics realizations, but the possibility is one that a designer of an optical computer may wish to explore.\footnote{The question has multiple facets: if the equations governing the optics and electronics dynamics were identical, one might still achieve an advantage of optics over electronics for some of the other reasons described in this article, such as bandwidth. However, one can also ask if the differences between the underlying equations lead to different behavior beyond a faster timescale resulting from higher bandwidth, or a larger system size resulting from larger spatial parallelism---in other words, differences beyond the other optics-vs-electronics distinctions drawn so far.}

\end{enumerate}

\begin{figure}[h]
  \includegraphics[width=0.45\textwidth]{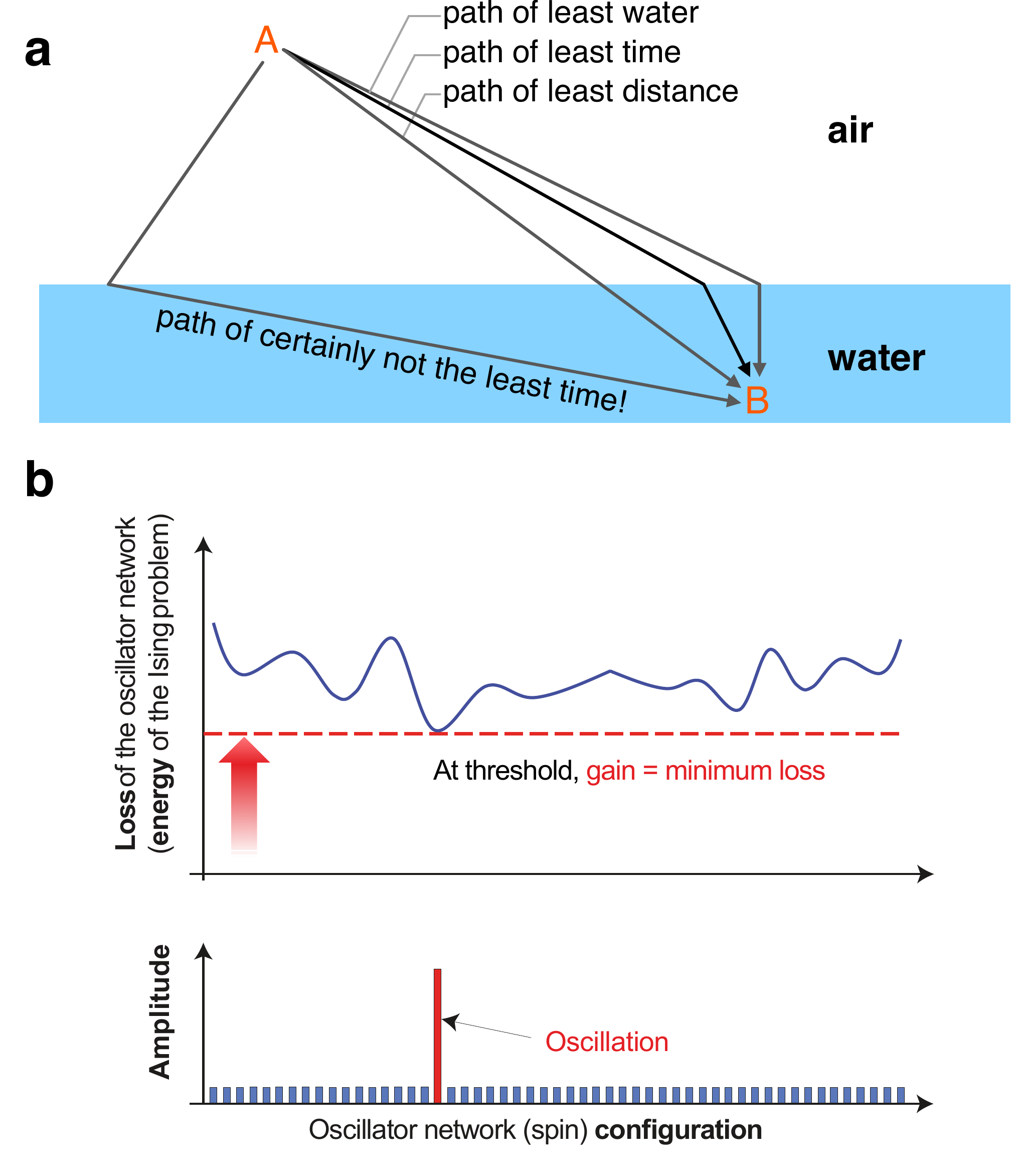}
  \caption{\label{fig:9_OptimizationPrinciples} \protect\circled{9} \textbf{Optimization principles.} \textbf{a}, The Principle of Least Time in optics. Light travels between starting point A and ending point B by taking the path of least time. A computational interpretation is that the light solves an optimization problem (of finding the path of least time), given the constraints of where the path starts and ends. \textbf{b}, A network of oscillators---which in optics could, for example, be optical parametric oscillators or laser oscillators---will in principle oscillate in the collective mode/configuration corresponding to the lowest loss if the gain is set to be equal to the minimum loss. {\color{gray} Panel \textbf{a} adapted from Ref.~\cite{feynman2006qed}, \textcopyright~Princeton University Press. Panel \textbf{b} adapted from Ref.~\cite{marandi2014network}, \textcopyright~Springer Nature.}}
\end{figure}

\newpage

\begin{enumerate}[label=\protect\circled{\arabic*}]
\setcounter{enumi}{9}

\item \textbf{The quantum nature of light is accessible at room temperature}:\footnote{In this paper, we do not consider quantum information processing \cite{nielsen2010quantum}; here, when we talk of operating in the quantum regime, we mean in the sense that light comprises photons and we are operating at such low powers that the quantum noise and discrete nature of the light is relevant to modeling the operation of the computer. The topic of using quantum phenomena such as entanglement to build quantum computers is exciting but beyond the scope of this paper; Ref.~\cite{dowling2003quantum} provides a helpful description delineating the first and second quantum revolutions, and it is only the former that we consider here.} it is possible to store and process information encoded with single optical-frequency photons, and it is possible to detect individual optical photons with low noise. This is in contrast to the situation at microwave frequencies, where thermal noise at room temperature rapidly swamps any information stored in single photons, and low-noise single-photon detection is not available.\footnote{The quantum nature of microwave photons is accessible at temperatures $\unsim \SI{10}{\milli\kelvin}$, but such cold temperatures are generally only achievable using a dilution refrigerator, which is bulky and expensive (in money and energy).}

For classical information processing, the fact that small numbers of photons can be manipulated and measured naturally leads to a potential reduction in energy cost versus if more photons were needed for reliable operation \cite{hamerly2019large,wang2022optical}. It is also possible to produce and measure squeezed states of light at room temperature \cite{andersen2016years}; the reduced noise in squeezed states could prove useful in classical information processing, for example for achieving higher numerical precision with a fixed energy budget (average number of photons).

The lack of a strong single-photon nonlinearity in optics, which is an advantage for communicating information without cross-talk but can be a disadvantage for processing information with small numbers of photons, can be circumvented using single-photon detection. The nonlinearity of the detection process itself is a feature one can use \cite{hamerly2019large,wetzstein2020inference,ma2023quantum}, but it is also possible to use photodetection to probabilistically induce nonlinear operations across multiple optical modes \cite{knill2001scheme}.\footnote{Ref.~\cite{knill2001scheme} develops and motivates probabilistic nonlinear operations for use in quantum computing, but these operations could potentially also be used for classical computing.}

\end{enumerate}

\begin{figure}[h]
  \includegraphics[width=0.5\textwidth]{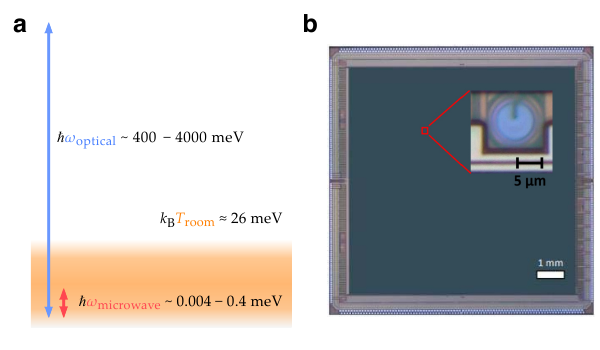}
  \caption{\label{fig:10_RoomTempQuantum} \protect\circled{10} \textbf{The quantum nature of light is accessible at room temperature.} \textbf{a}, The energy of optical photons is much higher than that of the thermal energy scale $k_\textrm{B} T$ at room temperature ($T_\textrm{room} \approx \SI{300}{\kelvin})$, whereas microwave photons have much lower energy than $k_\textrm{B} T_\textrm{room}$. Consequently thermal noise ``drowns out'' quantum effects of microwave signals at room temperature, but quantum effects in optical signals can be observed. \textbf{b}, An array of 250,000 single-photon detectors, which is sensitive to light at visible wavelengths and operates at room temperature. {\color{gray} Part \textbf{b} reproduced from Ref.~\cite{wayne2022500}, \textcopyright~IEEE.}}
\end{figure}

\newpage

\begin{enumerate}[label=\protect\circled{\arabic*}]
\setcounter{enumi}{10}

\item \textbf{Wave physics}\footnote{This point could have been presented as just a part of \textit{The quantum nature of light is accessible at room temperature}, since the wave--particle duality and the wave behavior of both photons and of electrons is part of quantum physics, but we have elevated it to being its own point because the wave nature of electrons being difficult to observe and exploit is not just due to cryogenic temperatures being required---on-chip electron coherence lengths are also much more dependent on the properties of the material host than on-chip photon coherence lengths.}: it is easy to observe the wave nature of individual photons---observing interference of single photons in a Mach-Zehnder interferometer is an undergraduate lab experiment \cite{pearson2010hands}, and photon coherence is well-preserved in on-chip photonic processors \cite{carolan2015universal}---but it is difficult to observe the wave nature of individual electrons\footnote{Even in advanced on-chip electron-transport experiments, the electron coherence length is less than $\unsim \SI{250}{\micro\meter}$, with values between $1$ and $\SI{20}{\micro\meter}$ \cite{duprez2019macroscopic} more typical, and only at cryogenic temperatures.}.

While this is true, a counterpoint is that even though the wave nature of \textit{individual electrons} is impractical to observe, wave phenomena of \textit{microwave signals in electronics} can easily be observed and exploited for computation \cite{lee2003design}. However, these are not wave phenomena of single electrons, but rather of signals that comprise many microwave photons. A key engineering consequence of this distinction is that electronic microwave signals have long wavelengths (e.g., GHz signals have centimeter-scale wavelengths) and this dramatically limits the possible spatial parallelism relative to the parallelism possible with optical-frequency photonic signals---leading to a potential advantage of optics over electronics (and in particular, microwaves).\footnote{A completely different kind of microwave signal can also be created and used for computation: an acoustic wave at microwave frequencies \cite{safavi2019controlling}. These waves can have short wavelengths despite their low frequencies, but at the cost of propagating at vastly slower speeds than photonic signals---the speed of sound instead of the speed of light---which is a disadvantage for computing with them.}

\end{enumerate}

\begin{figure}[h]
  \includegraphics[width=0.5\textwidth]{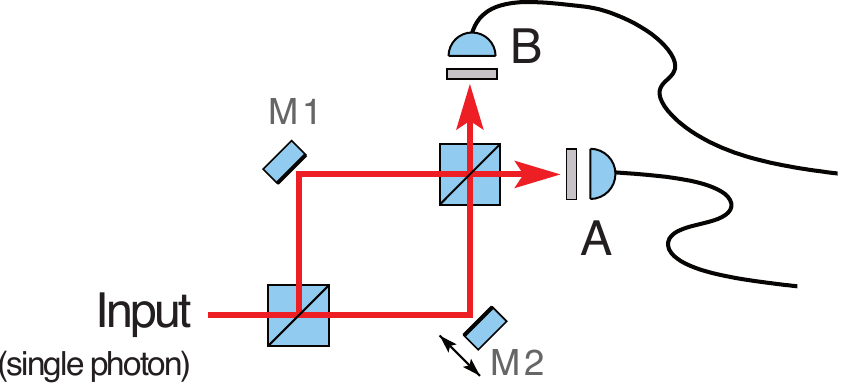}
  \caption{\label{fig:11_Waves} \protect\circled{11} \textbf{Wave physics.} Interference can be observed in a Mach-Zehnder interferometer with only a single photon input at a time. This schematic is from an undergraduate-laboratory experiment using just a few commercial optical components, highlighting the relative ease of observing wave phenonmena at the single-photon level with optics. (The counts at Photodetector A will oscillate as a function of the position of Mirror M2, which controls a phase difference between the upper and lower arms of the interferometer.) {\color{gray} Adapted from Ref.~\cite{pearson2010hands}, \textcopyright~AAPT.}}
\end{figure}

\newpage

\begin{enumerate}[label=\protect\circled{\arabic*}]
\setcounter{enumi}{11}

\item \textbf{\st{The speed of light is fast}}: the speed of light is often brought up as a reason for how optical computing will obtain a large speed advantage over electronic computers, but this is misleading because both optical and electrical signals can travel at roughly the same speed: in vacuum, light (and microwaves) travel at speed $c$; in silicon-photonic waveguides, light travels at speed $\unsim 0.4 c$ \cite{dwivedi2015experimental}; in wires on printed circuit boards, signals can travel at speed $\unsim 0.43 c$ \cite[Chapter~4]{rabaey2002digital}; and in CMOS electronic circuits, signals can travel at speed $\unsim 0.2 c$ \cite{miller2017attojoule}.\footnote{With careful design, speeds $\approx 0.5 c$ are possible in CMOS wires \cite{jose2006pulsed}.}

There is a mere $5 \times$ difference between the speed of light in vacuum and the speed of signal propagation in wires in CMOS electronic processors, so the speed of light is not a key distinction of optics. The notion of ``computing at the speed of light'' \cite{wetzstein2020inference} is more useful to think of as a \textit{goal} for an optical computer, rather than a \textit{cause} of advantage. The speed of light provides a physical limit on how fast a computer can operate \cite{bremermann1967quantum} and one framing of the optical computer engineer's goal is to design a computer that leverages the benefits of optics (\circled{1}--\circled{11}) to reach this limit for a particular computing task, in as small a volume as possible, so that the total time for a computation is as small as possible.\footnote{This framing implicitly makes the goal about the \textit{latency} of the computer (how long does it take for the answer to be output from the time the input is provided?)---which can be important, especially in real-time-computing scenarios---but often we are instead interested in improving its throughput or energy efficiency. Optimizing for throughput may involve trying to maximize the number of computing operations performed in parallel, and optimizing for energy efficiency may involve minimizing the dissipation in the system, neither of which have much to do with ensuring that the computer's latency saturates the bound set by the speed of light. ``Computing at the speed of light'' is not only a goal rather than a cause, but it is just one of several possible goals for an optical computer.}

\end{enumerate}

\newpage

Some of the features listed above are interrelated, and some of them even have a common physical root but are listed separately because the root leads to multiple features of light or has multiple consequences for computing. For example, the large bandwidth of optics (\circled{1}) relies on the large carrier frequency $\omega$ of optical signals. The wavelength of light $\lambda$ is directly connected with its frequency $\omega$: $\lambda$ is proportional to $1/\omega$, so the large values of $\omega$ for light make it possible to achieve large spatial parallelism ($\circled{2}$) and to observe and exploit wave physics in small volumes ($\circled{11}$). The fact that optical photons have a large energy $\hbar \omega$ relative to thermal energy $k_\textrm{B} T$ at room temperature $T \approx 300\textrm{~K}$ \footnote{$k_\textrm{B}$ is Boltzmann's constant.} is directly responsible for the quantum nature of light being accessible at room temperature (\circled{10} and \circled{11}). Low-dissipation dynamics (\circled{3}) and transmission of information with optics (\circled{4}) are also connected with the short wavelength $\lambda$ for optical photons, which allows tight waveguided confinement with nearly lossless dielectrics rather than with metals.\footnote{Microwave signals propagating through media such as metal coaxial cables or metal on-chip transmission lines suffer from substantial loss unless the metal is superconducting.} So all six of these features are connected by the fact that $\omega$ is large, since multiple aspects of optical physics are influenced by the value taken by $\omega$.

Not all of these features are equally important for obtaining advantage in optical computing but they are also not presented in order of importance, partially because determining such an order would require knowing what ingredients future optical computers will ultimately most heavily rely on. Nevertheless, in the next section we will discuss how these features may be used and opine on which ones are most likely to be critical.

\newpage

\section{Discussion}
\label{sec:discussion}

\textbf{How might optical computers beat electronic computers?} We will describe some strategies for the design of optical computers that may enable them to have an advantage over electronic computers.

There are three main metrics of computing performance for which we might aim to achieve an advantage: \textbf{latency}, \textbf{throughput}, and \textbf{energy efficiency}. Which of the three (or which combination) one targets in designing an optical computer depends on the user's goals, but there are arguments for how optics could enable advantage in all three of these metrics.\footnote{There are several other metrics of computers that are important, such as \textit{size}, \textit{robustness}, \textit{cost}, \textit{security} (susceptibility to hacking), and \textit{accuracy}. We don't have any reason to believe that an optical computer could deliver superior \textit{accuracy}, for example, than all possible electronic computers, so accuracy is not a metric we expect an optical advantage for, but instead we will typically aim to achieve an advantage in latency, throughput, and/or energy efficiency for a specified accuracy. Similarly the other metrics (size, etc.) provide other constraints that an optical computer must satisfy to be competitive for some particular use case.}

We now briefly describe these metrics using a particular computing example: machine-learning inference, and even more specifically, face recognition in an image. \textit{Latency} (also called \textit{delay}) refers to the time it takes for the computer to make a prediction of the name of the person in an image from the moment the computer is given the input image. \textit{Throughput} refers to how many inferences can be performed per second; for face recognition in images, a throughput metric is images processed per second.\footnote{Note that in general, (1/Latency) $\neq$ Throughput; by pipelining \cite{hennessy2017computer}, throughput can be much higher than the inverse of latency. As an intuitive example of this, consider a factory producing cars using an assembly line (pipeline): from start to finish, it might take the factory one day to manufacture a car (latency), but the total number of cars manufactured per day could be hundreds (throughput).} \textit{Energy efficiency} refers to how much energy is used by the computer to complete a single inference computation with a specified accuracy; for face recognition in images, an energy-efficiency metric is joules per image processed.

There may be trade-offs when optimizing for these three metrics, so it is important to decide before starting the design of a computer what one's goals are. For example, while minimizing \textit{latency} is sometimes the main goal (e.g., in high-frequency trading \cite{kao2022fpga}), often improving the throughput of a processor or its energy efficiency is the more important goal---and in many cases the goal will involve all three metrics, such as maximizing throughput and energy efficiency, subject to the constraint that the latency meets a particular target (e.g., in neural-network inference \cite{pope2022efficiently}, where in many applications---such as language translation---we may require the latency to be $<1$ second).

Despite the fact that there will typically be trade-offs in the optimization of computer performance metrics (e.g., between latency and throughput), the following strategies should help in designing a computer that optimizes any combination of latency, throughput, and energy efficiency:

\begin{enumerate}
    \item \textbf{Avoid or mitigate input/output bottlenecks and overheads.} Optical computers generally do not operate entirely with optics: typically some inputs to the computer originate in electronics, and/or the output from the computer is ultimately electronic. For example, if an optical processor is used for determining if there is a pedestrian walking in front of a self-driving car, the output needs to be electronic so that it can be input to the control systems in the car, which can use the information to actuate the brakes. If the processor uses a neural network, the trained parameters for the neural network may well be stored in electronic memory and need to be input to the processor in some way. Unfortunately the interfaces between optics and electronics can cause major bottlenecks in speed and be a major source of energy usage by a processor. For an optical processor to offer an advantage over electronic processors---in any of latency, throughput, or energy efficiency---the processor architecture needs to be designed to minimize the negative impact of transduction between optical and electrical signals, and the conversion between analog and digital signals.
    
    To illustrate some of the challenges that can arise from optics-electronics interfaces, imagine an optical processor that intrinsically has a processing bandwidth of $\SI{100}{\tera\hertz}$ (\circled{1}). If data can only be input to the processor at a rate of $\SI{10}{\giga\hertz}$, limited by, for example, the bandwidth of electro-optic modulators and digital-to-analog converters, then without careful design, the intrinsic bandwidth benefit of the optical system---which could have led to improved latency and/or improved throughput---may go to waste. Similarly, while an optical processor can be designed to perform computation on optical signals nearly dissipationlessly, there is an energy cost to optical/electrical transduction and analog/digital conversion for getting electronic data into and out of the optical processor, and these costs may be so large that they don't just dominate the total energy cost of the optical processor, but make the energy cost so high that the processor is less energy efficient than an all-electronic processor.

    A crucial mitigation strategy is to \textbf{re-use data} that is input as much as possible---once you have paid both the time and energy penalty for sending electronic data into an optical processor, you would like to extract as much benefit as possible from that data. This applies both to data converted into optical signals and to data that may remain as electrical signals but that nevertheless has time and energy costs to be input to the processor. Re-use of optical signals can be enabled by various forms of optical memory \cite{alexoudi2020optical}, as well as by copying via fanout (\circled{7})\footnote{Consequently an optical-computer designer is usually motivated to make the fan-out factor be as large as possible. In an optical matrix-vector multiplier, fanning out $10^3$ or more copies of the input vector is desirable, and likely necessary to achieve a substantial advantage over electronics.}. As an example of the re-use of electrical control signals, optical processors performing neural-network inference (as opposed to training) can load the neural-network weights into phase shifters that consume either little or no static power \cite{shen2017deep,wetzstein2020inference} and then use those weights many times by performing many inference computations with them (for example, by batching individual inferences \cite{hamerly2019large}). This allows both the time and energy costs of loading the weights to be amortized. Another example of data re-use in photonic neural-network processors is in convolutional neural networks: the same convolutional kernel can be applied to many different subsets of the input data, so the kernel weights can---at least conceptually---be loaded once and used many times \cite{wetzstein2020inference,xu2021tops,feldmann2021parallel}.
    
    A general design principle is that it is---all else held equal---\textbf{better to perform more computations per bit of input data}. This is essentially the concept of maximizing \textit{arithmetic intensity} in conventional computer architecture \cite{hennessy2017computer}. Data re-use is one way to achieve this, but an important complementary conceptual approach is to choose computational tasks such that the optical processor for that task performs computations whose complexity scales rapidly with the input data size. For example, a computation on input data of size $N$ that requires only $O(N)$ operations is less attractive than one that needs $O(N^2)$ operations, and a computation requiring $O(N^3)$ operations is even better. The cost in time and energy of inputting data of size $N$ is generally $O(N)$, so if the computation performed by the optical system has complexity $O(N^2)$ (and we assume that, through a combination of \circled{1}--\circled{11}, the cost of this computation in optics is far lower than it is in electronics) then there exists some threshold size such that for any $N$ larger than the threshold, the costs of loading the data can be compensated for by the benefits of doing the $O(N^2)$ computations optically---leading the optical computer to outperform electronic computers even when the data-transfer costs are considered. A key practical fact is that for current speed and energy numbers for CMOS electronics, it seems likely that optical processors will need to support very large values of $N$ (e.g., $N>10^4$) to reach the crossover point where they start delivering a throughput or energy-efficiency advantage for computations based on matrix-vector multiplication (which is an $O(N^2)$ computation, for square matrices) \cite{anderson2023optical}. This fact motivates both scaling optical matrix-vector-multiplication processors to large sizes and designing optical processors with computations that have complexity greater than $O(N^2)$. From this perspective, combinatorial optimization such as Ising solving \cite{mohseni2022ising} is an attractive problem for optical computing because the computing effort is generally expected to scale exponentially, i.e., as $O(2^N)$, with respect to the number $N$ of variables being optimized, and also with the amount of data required to specify the optimization problem\footnote{For example, an $N$-spin Ising problem is specified by $O(N^2)$ numbers.}.

    Because the cost of loading data is generally larger for optical processors than for electronic processors\footnote{When an optical processor loads data from electronic memory, there is not only a cost for the memory access---which an electronic processor would also have had to pay---but there is a cost for transducing the data from an electrical to an optical signal, and potentially also a digital-to-analog conversion involved, which also has a cost.}, there is a strong motivation to choose algorithms for optical processors that have higher intrinsic data re-use or higher algorithmic complexity. This kind of hardware-software co-design can lead to considerable improvements when compared with fixing the algorithm based on what works well on current electronic processors and trying to forcibly design an optical processor to work in the same way.

    While minimizing and compensating for the costs of loading input data is crucial, it is also important to avoid having the output of data be too costly in time or energy. It is similarly beneficial to \textbf{minimize how much data needs to be output}, by doing as much of the computation and data reduction within the optical processor as possible. This design principle motivates choosing algorithms that require a large amount of computation relative to the size of the output. As an example, this is typically true in machine-learning inference---where for the overall computation the answer may be just a few tens of bits, outputting the predicted class of the input data.

    \item \textbf{Don't try to directly take on digital electronic processors at their own game.} Arguably the biggest challenge in building optical processors that surpass electronic processors in throughput or energy efficiency is overcoming the limiting performance of electronics-to-optics and optics-to-electronics conversion technology. If we start with data in electronics---as is most typically the case---and want our computed answers to end up in electronics---as is also most often the case---then we have little choice but to apply the strategies above and hope to be able to amortize the input/output costs. However, given how large state-of-the-art CMOS electronic processors are and that they have a home-ground advantage in working on data that is already in electronics, it seems likely that modern optical processors won't first gain an advantage as drop-in replacement accelerators in conventional electronic processing workflows. Instead, we can \textbf{target applications where the inputs and/or outputs are naturally optical}---and in this way eliminate the conversion costs.
    
    Machine-learning applications where the input is conventionally an image from a camera is an example \cite{wetzstein2020inference,ashtiani2022chip,wang2023image}: one can replace the camera and subsequent electronic neural network with an optical neural network that directly processes the scene in front of it, e.g., in self-driving cars \cite{rodrigues2021weighing}, microscopy \cite{wang2023image}, or spectroscopy. It is not necessary to replace all the electronic image-processing computation with optics if the output is ultimately going to be electronic anyway---one can adopt the strategy of using optics to pre-process the optical image data \cite{chang2018hybrid,colburn2019optical}, intelligently encoding it so that the output conversion from optics to electronics has much lower bandwidth than naively digitizing the images to begin with, which could lead to benefits in latency, throughput, and energy efficiency \cite{wang2023image}.
    
    While image processing enables the elimination of the input conversion stage because the input can be directly optical, applications where both the input and the output are optical may be even more promising for immediate attack. Optical communications have inputs and outputs that are both optical, but current approaches involve a number of stages at which optical signals are converted to electrical signals for electronic processing, and then converted back to the optical domain. This makes optical communications signal processing a natural target for all-optical signal processing, which could reduce latency, increase throughput, and improve energy efficiency \cite{minzioni2019roadmap,huang2021silicon,huang2022prospects,chen2022photonic}.

    Many neural-network models have become large enough that they can no longer practically be run on a single electronic processor, which has motivated the design of optical interconnects specifically for neural-network processing \cite{ghobadi2022emerging}. This trend provides another motivation for neural-network processing as an application for optical processors: if the electronic-processor competition needs to pay the relatively high energy costs of conversion between optics and electronics too, then these conversion costs are at least not an exclusive disadvantage of using optical processors. One can think of a single processor in an optically-interconnected datacenter for performing neural-network processing as a system whose inputs and outputs are both optical---so from this perspective, it is a promising candidate to try replace with an optical processor.
    
    \item \textbf{Combine multiple optical features to try gain an advantage.} This might sound trite, but it is important---any optical processor that has an advantage over the best equivalent electronic processors will most likely need to take advantage of not just one of the features of optics (\circled{1}---\circled{11}), but will need to carefully combine several of them. For example, just taking advantage of the large bandwidth of optics (\circled{1}) in a single spatial mode---even if we ignore for now input/output bottlenecks---is probably not sufficient to enable a throughput benefit since electronic processors compensate for lower bandwidth with enormous spatial parallelism (having on the order of $10^{11}$ transistors in modern chips). Similarly only relying on spatial parallelism (\circled{2}) will likely also be insufficient: while the spatial parallelism of optics is considerable, especially in three-dimensional systems, the the spatial parallelism of transistors is typically even more impressive.\footnote{Optical multiplication of vectors by random matrices is an exception where the spatial parallelism is so large that even very low bandwidth doesn't prevent the system from having higher throughput that electronic processors \cite{saade2016random}. Even in this case though, more than one property of optics is being used: for example, not just spatial parallelism (\circled{2}), but also nearly dissipationless dynamics (\circled{3}).} However, if one can combine the bandwidth and spatial-parallelism features of optics in a single system, then there is potential to surpass electronics. For example, imagine being able to process data in $10^7$ spatial modes in parallel at a clock rate of $\SI{10}{\tera\hertz}$, or processing data in parallel in $10^7$ spatial modes, each with $10^7$ frequency modes---in other words, $10^{14}$ parallel spatio-frequency modes.\footnote{Where did the numbers $10^7$ and $10^7$ come from? They were chosen somewhat arbitrarily but as believably practical, since, for example, we already have technology---spatial light modulators---for manipulating $10^7$ spatial modes. We could have even higher numbers of spatial and frequency modes though---this was an example, not a bound.} Although it is far from a solved problem how to fully take advantage of the combination of bandwidth and spatial parallelism afforded by optics, when combined with the fact that operations can be performed nearly dissipationlessly in optics (\circled{3}), there is great potential for optics to outperform electronics.

    Accurately predicting the future of technology is difficult, but it seems reasonable to hypothesize that of the 11 features explored in this paper, bandwidth (\circled{1}), spatial parallelism (\circled{2}), and nearly dissipationless dynamics (\circled{3}) are most likely to play a key role in any future optical processor that does deliver an overall advantage (in latency, throughput, or energy efficiency). However, many of the other features (\circled{4} -- \circled{11}) may very well end up playing important roles too, so should not be ignored---but they will probably need to be combined with one of the ``big three'' (\circled{1} -- \circled{3}) for a processor using them to achieve an overall advantage over electronics.
    
    Many of the demonstrations of optical processors to date have shown a proof of principle of the use of some feature of optics for computing in a way that could lead to an advantage, but with a system that doesn't suitably leverage some of the other available features, ultimately leading to a prototype that is inferior to current electronic processors. An example of this from my own group is Ref.~\cite{wang2022optical}, which uses spatial parallelism (\circled{3}) to realize $>500,000$ scalar multiplications per pass of light through a free-space optical processor, but the prototype is extremely limited in bandwidth (\circled{1}) due to the speed limits of the input and output stages, leading to performance that is ultimately many orders of magnitude worse than an electronic processor. In this work we were not expecting to beat an electronic processor but rather were aiming to demonstrate how few photons are needed for matrix-vector multiplication in optical neural networks; nevertheless, to advance this proof-of-principle system to be competitive with electronics would require dramatically increasing the system bandwidth.\footnote{Besides spatial parallelism (\circled{3}), the optical processor presented in Ref.~\cite{wang2022optical} also used some other features of optics, such as nearly dissipationless dynamics (\circled{3})---without which the ultra-low optical energy usage demonstrated would not have been possible, and optical fan-in (\circled{7}).}
\end{enumerate}

My opinion is that the most likely route to building an optical processor that delivers a large advantage over electronic processors in throughput or energy efficiency (or both) in the near term is by constructing a free-space optical matrix-vector-multiplier that takes advantage of large spatial parallelism (\circled{2}) and nearly dissipationless dynamics (\circled{3}) \cite{wetzstein2020inference}. With a vector dimension of $N \approx 10^4$ and a matrix size of $N \times N$, it seems promising that one can achieve an advantage provided that the system can be operated at a rate of one matrix-vector multiplication per nanosecond and the surrounding electronics for input and output operate with state-of-the-art energy efficiency \cite{wang2022optical,anderson2023optical}. This will require careful optical and electronic engineering to realize---it is a major engineering undertaking whose difficulty should not be underplayed---but is all based on existing technology components that can in principle be appropriately scaled. I find this candidate architecture the most promising in the near term largely because it has been well-studied and many of the necessary building blocks are fairly advanced. An optical matrix-vector-multiplier whose inputs are optical, such as when it is used as a preprocessor for visual scenes \cite{wang2023image}, would have a lower bar to deliver an advantage over electronic solutions, so I expect that if an optical matrix-vector-multiplier does outperform an electronic processor it will probably first be for an application involving optical inputs. However, I certainly don't want to give the impression that I think a free-space spatially multiplexed architecture is the only one worth pursuing. There are a multitude of other architectures \cite{wetzstein2020inference,shastri2021photonics}, including those based on photonic integrated circuits and on additionally taking advantage of the large bandwidth of optics (\circled{1}), that are appealing and very much worth pursuing.

When evaluating an optical-computing scheme---especially one relying on Features \circled{9}--\circled{11}, for which it is often unclear if the optical scheme is not just \textit{different} from a standard digital-electronics solution, but \textit{better}---it can be helpful to determine what the cost of simulating the scheme with a digital-electronic processor would be. For example, wave physics (\circled{11}) can be simulated by digital electronic processors\footnote{In the case of simple interference, this can be as easy as adding two complex numbers.}, so when seeking an advantage for optics from wave phenomena, one needs to consider the cost of equivalent digital electronic approaches, and depending on the wave phenomena being exploited, the digital approaches may be competitive or outright superior. Some intuition for how wave physics in optics could be exploited to give an advantage over digital simulation of wave physics is that in the single- or few-photon regime, the optical energy used could be very small, and relates to the feature of nearly dissipationless dynamics in optics (\circled{3}). As another example, least-power-dissipation principles (\circled{9}) can be used to realize Ising optimizers from networks of coupled optical oscillators \cite{mohseni2022ising}, but simulating the equations of motion of the network on a digital-electronic computer can yield the same behavior as a physical, optical implementation, so the intrinsic least-power-dissipation phenomenon doesn't automatically give rise to a computing benefit. Instead one also needs to leverage other benefits of optics, such as parallelism and low dissipation.

We conclude by summarizing some of the \textbf{major outstanding challenges} that, if addressed, would move us substantially closer to realizing practically useful optical computers:

\begin{itemize}
    \item \textbf{Optical-processor architecture design.} There is a major challenge to design architectures of optical processor that most effectively use the features of optics to gain an advantage. It is not obvious that the existing optical-processor architectures (using free space or integrated photonics)---some of which are decades old \cite{ambs2010optical}---are optimal, and there is an opportunity to invent refined or completely new designs to meet this challenge.
    
    \item \textbf{Applications.} We need to find good applications to target with optical processors. Since one of the major roadblocks to achieving advantage with optical computing are issues associated with input/output, we want to find valuable applications where we can avoid or mitigate input/output bottlenecks and costs. For example, it has proven very difficult to build an optical matrix-vector multiplier at a scale ($N$) at which the input/output costs can be sufficiently amortized, even though an optical matrix-vector multiplier can perform $O(N^2)$ operations with input/output costs of just $O(N)$.\footnote{For simplicitly, the expressions given here assume a square matrix of size $N \times N$.} Given that even matrix-vector multiplication, with its $O(N^2)$ complexity, does not have a high enough ratio of computation to input data, it would be helpful to find useful subroutines, algorithms, or applications that have higher complexity than $O(N^2)$ for input and output data sizes $\unsim N$.

    An additional direction is to find applications that could benefit from other aspects of optical computing besides potential performance advantages. For example, direct optical processing of visual scenes could give a privacy advantage: an electronic processor of images captured by a camera that stores the images in memory could be hacked, but an optical processor that directly processes what is ``sees'' and never converts the full incoming images to electronic format could be a lot harder to maliciously copy images from.
    
    \item \textbf{Nonlinearity.} Nonlinearity is crucial in many computations and a low-energy, fast, small-footprint, reliably manufacturable nonlinearity would be a useful building block. The nonlinearity need not necessarily be all-optical---optoelectronic nonlinearity can also be useful \cite{williamson2019reprogrammable}, although generally one can hope to benefit from higher bandwidths and possibly lower energy consumption in all-optical nonlinearities \cite{guo2022femtojoule}. A fast, few-photon nonlinearity capable of attojoule switching has recently been demonstrated \cite{zasedatelev2021single}; one important direction is in scalably manufacturing the nonlinearities that have already been established.

    \item \textbf{Cascadability.} In many computations---for example, in deep neural networks---the input data is fed not through just one function but a sequence of functions. An optical implementation of the computation then often involves passing an optical signal either through the same optical setup multiple times or through multiple different optical setups (or both). This requires being able to cascade optical processes in time or space. Three of the challenges\footnote{The first listed challenge---that of optical attenuation---is nearly universal in optical processors. The other two challenges listed are specific to optical-computing schemes using particular implementations of optical nonlinearity.} that can arise in cascading optical processing stages are attenuation of the optical signal due to optical loss, effective attentuation of the optical signal due to weakness in optical nonlinearity\footnote{Because optical nonlinearity is generally weak \cite{boyd2020nonlinear}, less than 100\% of the light input to a nonlinear stage will generally be acted on nonlinearly. If the relevant part of the optical signal output from a nonlinear stage is the light that experienced the nonlinear transformation and the part that was not affected is discarded---either explicitly, or implicitly by not taking meaningful part in later stages of the computation---then effectively there is an attenuation of the signal that is fundamental but is unrelated to optical loss, in that it would occur even if the optical system were lossless.}, and nonlinear-optical processes generating output light that is at wavelengths incompatible with being input to the next optical stage\footnote{For example, directly cascading many second-harmonic-generation processes is infeasible, since the optical signal's frequency is \textit{doubled} at each stage, so after just a few stages one reaches wavelengths that are beyond the optical spectrum and are impractical to use.}. Designing suitably cascadable systems can be approached in multiple ways: for example, at the level of processor architecture, one may opt to insert gain into the system to compensate for the signal attenuation---which leads to further architectural and system-design decisions about the type of gain (purely optical or optoelectronic, in which case the gain is essentially provided electronically by transistors\footnote{Using optoelectronic gain to enable cascadability---and often also serving the dual purpose of providing nonlinearity---is one of the most common architectural choices in optical-neural-network prototypes \cite{shastri2021photonics}.}), and its required speed, preservation of information encoded in the optical spectrum, and so on, as well as new engineering challenges in realizing suitable gain components. One may also approach cascadability challenges at the component or physical-implementation level, seeking to realize lower-loss optical systems, or materials with higher nonlinear coefficients.
    
    \item \textbf{3D design and manufacturing.} Spatial parallelism (\circled{2}) can be massively enhanced by using a third dimension, and if the dissipation is kept low (\circled{3}), this provides a path to advantage over electronics. Separately, enabling long-range coupling between modes by using a third dimension (and advantages from Features \circled{4} -- \circled{6}) can also bring benefits \cite{dinc2020optical,boahen2022dendrocentric}. The key question here is \textit{how} to engineer and fabricate programmable, large-scale, possibly dense, three-dimensional processors \cite{psaltis1990holography,morris2012dynamic,dinc2020optical,moughames2020three}.
    
    \item \textbf{Energy costs for electronic and optoelectronic components.} The energy cost of optical processors is typically dominated by the energy costs of the electronic parts of the computer (for example, in an analysis of optical neural networks running large Transformer models, the optical energy used accounts for $<1\%$ of the total energy cost \cite{anderson2023optical}; see also Refs.~\cite{hamerly2019large,tait2022quantifying}). Many optical-computing schemes could benefit from---and to deliver advantage, may even require---the availability of large arrays of high-speed, low-power, and low-cost detectors, analog-to-digital converters, modulators, and digital-to-analog converters. Increasing the energy efficiency of these components is an important challenge.
    
    \item \textbf{Scale.} Most optical-computing schemes rely on parallelism---be it from frequency or time multiplexing (\circled{1}), or spatial multiplexing (\circled{2}), or a combination---for part of how they will achieve an advantage over electronics. However, throughput and energy-efficiency advantages typically only materialize when the system size (i.e., the number of parallel operations) is very large \cite{anderson2023optical}.\footnote{The situation for \textit{latency}, as opposed to throughput or energy-efficiency, advantages is more subtle in that it is more application-dependent: if an application requires a certain amount of highly parallelizable computation (such as matrix-vector multiplication) to be performed in as little time as possible, so long as an optical processor is large enough to perform all that computation in parallel, it is big enough and won't necessarily benefit from larger scale (from the perspective of latency). A latency advantage could then arise from how the system is designed to minimize the time it takes to get the data into and out of the constituent parallel-processing units. But on the other hand, an optical processor could also deliver a latency advantage that is directly attributable to its scale: if it has parallelism far beyond that of an electronic processor it may achieve a throughput advantage that then will typically give a latency advantage as a side benefit for large tasks where an electronic processor would need to perform the computation in multiple stages in series on account of the task being larger than the electronic processor's parallel-processing capacity.}
    
    For example, we would like optical matrix-vector multipliers to be large enough to amortize the energy costs of loading the input vector and reading out the output vector. We would also like them to be large enough to be able to compete in throughput with electronic processors, which can perform $>10^6$ 8-bit-precision scalar multiplications per nanosecond \cite{nvidia2022nvidia}---so if vectors are input at a rate of $\SI{1}{\giga\hertz}$, we would like the optical processor to also be able to perform $>10^6$ scalar multiplications in parallel. However, in optical matrix-vector multipliers made from arrays of Mach-Zehnder interferometers \cite{wetzstein2020inference}, even a state-of-the-art commercial prototype with a $64 \times 64$ array \cite{ramey2020silicon} does $>100 \times$ fewer parallel operations than seems necessary to compete in throughput with state-of-the-art electronics solutions. A major challenge is how to scale arrays of size $64 \times 64$ to something much larger, like $1000 \times 1000$, which would put them roughly on par with the degree of parallelism in a single state-of-the-art electronic chip \cite{nvidia2022nvidia}, or $10^4 \times 10^4$, which would then be in the regime where a substantial throughput advantage could be achieved provided the system were clocked at a comparable rate to electronics (i.e., at $\unsim \SI{1}{\giga\hertz}$). How can Mach-Zehnder-interferometer arrays be scaled from sizes $\unsim 64 \times 64$ to sizes $\unsim 10^4 \times 10^4$? This is a major challenge for the community working on this approach.

    The challenge of scaling to achieve a far greater degree of parallelism than current prototypes is certainly not unique to optical matrix-vector multipliers or Mach-Zehnder-interferometer arrays---most optical-computing schemes face a major scaling challenge for them to be able to deliver a practical advantage. In some cases, we don't even have a solid practical roadmap for how to scale yet: for example, what is a feasible way to scale a scheme that combines spatial and frequency multiplexing (such as that in Ref.~\cite{feldmann2021parallel}, using 16 spatial and 4 frequency degrees of freedom) to a point where it can achieve advantage? There is the potential for very large numbers of both spatial and frequency modes to be harnessed to perform parallel computations (e.g., $>10^{14}$ spatio-frequency modes being operated on in parallel), but how can we reach this scale for a concrete scheme that performs useful computation?

    \item \textbf{Robustness, reliability, and fabrication variation.} While many optical components, such as those appearing in consumer-electronics devices like cellphones and in optical-fiber-communications systems, are generally very reliable, there are many optical technologies that are being considered for use in optical computers that present challenges in robustness (e.g., how well they can perform in the presence of environmental perturbations such as temperature changes or mechanical vibrations), reliability (e.g., how likely they are to keep functioning correctly under normal operation conditions), and fabrication variation (e.g., how much fabricated devices will differ in specifications from their designed values). For example, many optical phase-change-memory technologies have stringent limits on how many times they can be switched, and it is desirable for these limits to be raised \cite{zhang2021myths,martin2022endurance}. As another example, in integrated photonics, Mach-Zehnder interferometers typically suffer from the constituent splitters have small deviations from the ideal splitting ratio due to variations in fabrication; one research direction is to improve the fabrication processes, and another is to construct designs that can compensate for these fabrication errors \cite{hamerly2022asymptotically}. Generally for each photonic technology platform that might be used in an optical computer, there are open problems in how to stabilize them---passively or actively.
    
    \item \textbf{Storage.} To avoid the costs of converting between electronics and optics, and to avoid the cost of electronic memory accesses (which are a dominant cost even in electronic computing \cite{horowitz2014computings}), we would often like to be able to store data for use in optical processing. For example, in matrix-vector multipliers, we typically want to be able to store matrices with as low energy cost as possible for maintaining the storage, but in a way that the matrix can be updated on demand many times, at reasonably high accuracy (e.g., 8 bits), and also with relatively low energy cost \cite{bogaerts2020programmable,zhang2021myths}. In some applications or architectures, it is advantageous to be able to store optical signals (e.g., corresponding to intermediate calculation results) so that conversion from optics to electronics and then back to optics can be avoided. There is active study and much room for improvement in both these use cases of storage.
    
    \item \textbf{Pushing toward quantum limits.} Operating optical computers in a regime where the quantum nature of light cannot be ignored, e.g., by using ultra-low optical powers where signals comprise small numbers of photons and are measured by single-photon detectors, is a path toward minimizing optical energy consumption. Optical computers will inevitably involve some electronics, if only for control or readout, and it is often the electronics energy costs that dominate \cite{tait2022quantifying}, so it is only in some cases that there is strong benefit to minimizing the optical power used. Nevertheless, for these situations, there is much work to be done in both designing architectures and realizing practical devices that benefit from operating in the quantum regime \cite{mabuchi2011nonlinear,kerckhoff2011remnants,tezak2015coherent,shainline2017superconducting,ma2023quantum}.\footnote{In this paper, we do not consider quantum information processing \cite{nielsen2010quantum}; here, when we talk of operating in the quantum regime, we mean in the sense that light comprises photons and we are operating at such low powers that the quantum noise and discrete nature of the light is relevant to modeling the operation of the computer. The topic of using quantum phenomena such as entanglement to build quantum computers is exciting but beyond the scope of this paper; Ref.~\cite{dowling2003quantum} provides a helpful description delineating the first and second quantum revolutions, and it is only the former that we consider here.}
\end{itemize}

Constructing an optical computer that beats an electronic computer in any metric is challenging given how advanced electronic processors are. However, the physics of optical computing gives promise that if optical computers are carefully engineered, for certain classes of tasks---especially those involving data that is already in an optical format or that has a very high ratio of computation to data---they may deliver orders-of-magnitude benefits in latency, throughput, or energy efficiency.

\section*{Acknowledgements}

I gratefully acknowledge many helpful conversations with colleagues including Daniel Brunner, Ryan Hamerly, Hideo Mabuchi, Arka Majumdar, Alireza Marandi, Edwin Ng, Tatsuhiro Onodera, Tianyu Wang, Logan Wright, and Yoshihisa Yamamoto; these conversations over several years have shaped my understanding of optical computing. I also gratefully acknowledge Sapan Agarwal for explanations about analog-electronic crossbars, and Bal Govind for discussions about electrical interconnects. I thank Maxwell Anderson, Tianyu Wang, and Fan Wu for providing detailed feedback on a draft of this manuscript. This work has been financially supported in part by the National Science Foundation (Award CCF-1918549), NTT Research, and a David and Lucile Packard Foundation Fellowship.

\bibliographystyle{mcmahonlab}
\bibliography{references}

\end{document}